%% file: RT_from_tadpoles_QED.tex
\documentclass[11pt,a4paper]{article}
\pdfoutput=1

\usepackage{jheppub}

\usepackage[english]{babel}
\usepackage{relsize}
\usepackage{mathrsfs}

\usepackage{amstext}  
\usepackage{amsfonts}
\usepackage{amsthm}
\usepackage{graphicx}
\usepackage{amsmath}
\usepackage{amsfonts}
\usepackage{mathtools}

\usepackage{array}                
\usepackage{xcolor} 

\usepackage{slashed}

\input{macrosO2L}

\graphicspath{{./figures/}}

\newcommand{\diawidth}{1.8cm}
\newcommand{\diaheight}{1.6cm}

\preprint{
\begin{flushright}
PSI-PR-20-01\\
ZU-TH 03/20\\
\end{flushright}
}

\title{\boldmath Rational Terms of UV Origin at Two Loops}

\author[a]{Stefano Pozzorini}
\author[a]{Hantian Zhang}
\author[b]{Max F. Zoller}

\affiliation[a]{Physik-Institut, Universit\"at Z\"urich, CH-8057 Z\"urich, Switzerland}
\affiliation[b]{Paul Scherrer Institut, Forschungsstrasse 111, CH-5232 Villigen PSI, Switzerland}

\emailAdd{pozzorin@physik.uzh.ch}
\emailAdd{hantian.zhang@physik.uzh.ch}
\emailAdd{max.zoller@psi.ch}

\abstract{
The advent of efficient numerical algorithms for the construction of
one-loop amplitudes has played a crucial role in the automation of NLO
calculations, and the development of similar algorithms at two loops is a
natural strategy for NNLO automation. 
Within a numerical framework the numerator of loop integrals is usually
constructed in four dimensions, and the missing rational terms, which arise
from the interplay of the $(\dendim-4)$-dimensional parts of the loop
numerator with $1/(\dendim-4)$ poles in $\dendim$ dimensions, are
reconstructed separately.
At one loop, such rational terms arise only from UV divergences and can be
restored through process-independent local counterterms.
In this paper we investigate the behaviour of rational terms of UV origin at two loops.
The main result is a general formula that combines the subtraction of UV
poles with the reconstruction of the associated rational parts
at two loops.
This formula has the same structure as the R-operation, and all poles and
rational parts are described through a finite set of 
process-independent local counterterms.
We also present a general formula for the calculation of all relevant 
two-loop rational counterterms in any renormalisable theory 
based on one-scale tadpole integrals.
As a first application, we derive  the full set of two-loop rational
counterterms for QED in the $R_{\xi}$-gauge.
}

\keywords{}
\arxivnumber{}

\begin{document}
\maketitle
\flushbottom

\section{Introduction}

Higher-order calculations of scattering amplitudes are usually performed
in $\dendim=4-2\eps$ dimensions~\cite{tHooft:1972tcz},
 where the ultraviolet (UV) and infrared
(IR) divergences of loop integrals assume the form of $1/\eps$ poles.%
\footnote{For a review of the
different variants of dimensional regularisation
see~\cite{Gnendiger:2017pys}.}
Upon subtraction of all UV and IR singularities, 
scattering amplitudes become finite in the 
limit $\eps\to 0$. Nonetheless they still contain non-vanishing
contributions stemming from the interplay of $1/\eps$ poles with 
the $(\dendim-4)$-dimensional
parts of loop integrands.
For non-trivial processes,
depending on the employed technique the explicit
calculation of such $(\dendim-4)$-dimensional 
parts
can be technically involved and CPU intensive.
For this reason, 
automated one-loop tools such as 
{\sc OpenLoops}~\cite{Buccioni:2019sur}, 
{\sc Recola}~\cite{Denner:2017wsf},
{\sc Helac-1Loop}~\cite{vanHameren:2009dr} and
{\sc MadLoop}~\cite{Hirschi:2011pa}
are based on numerical algorithms that
construct the numerators of loop integrals in four dimensions,
while keeping the denominators in $\dendim$ dimensions.
The missing contributions 
stemming from the $(D-4)$-dimensional parts of loop numerators
are easily reconstructed a posteriori
through insertions of 
process-independent rational 
counterterms~\cite{Ossola:2008xq,Draggiotis:2009yb,Garzelli:2009is,Pittau:2011qp}
into tree amplitudes.

More explicitly, let us consider the 
renormalised amplitude 
of a one-loop diagram $\gamma$,
\bea 
\label{eq:ren1lddim}
\textbf{R}\,\ampbar{1}{\gamma}{}{} &=&
\ampbar{1}{\gamma}{}{}+\deltaZ{1}{\gamma}{}{}\,,
\eea
where $\ampbar{1}{\gamma}{}{}$ denotes the unrenormalised amplitude in $\dendim$
dimensions, and $\deltaZ{1}{\gamma}{}{}$ is the corresponding UV counterterm. 
In this paper we focus on the contributions that arise 
when the loop-integrand numerator in $\dendim$ dimensions  is split into two parts,
\bea
\label{eq:numsplitting}
\bar\calN (\barq)
&=&
\calN (q)+
\tilde\calN (\barq),
\eea
where $q$ is the loop momentum, and symbols with and without a bar denote,
respectively, quantities in $\dendim$ and four dimensions, while $\tilde\calN(\bar
q)$ is the $(\dendim-4)$-dimensional part of the loop numerator.

At one loop, the interplay of $\ntilde(\bar q)$ with $1/\eps$
poles of IR type can generate finite terms at intermediate stages of the
calculations, but at the level of full Feynman
diagrams such terms 
cancel\footnote{More precisely, $\ntilde$-contributions of IR origin cancel
in regularisation schemes where the external degrees of freedom are kept in
four dimensions, such as in the 't~Hooft--Veltman
scheme~\cite{tHooft:1972tcz}.}~\cite{Bredenstein:2008zb}.
Thus $\ntilde$-contributions arise only from divergences of UV type.
This makes it possible to cast the 
renormalised amplitude \refeq{eq:ren1lddim} in the form
\bea 
\textbf{R}\,\ampbar{1}{\gamma}{}{} &=& \amp{1}{\gamma}{}{} + \deltaZ{1}{\gamma}{}{} + \ratamp{1}{\gamma}{}{}
\,,
\label{eq:ren1l4dim}
\eea
where $\amp{1}{\gamma}{}{}$ is the unrenormalised amplitude
with numerator $\calN(q)$ in four dimensions, 
$\deltaZ{1}{\gamma}{}{}$ is the usual 
$\msbar$ counterterm, 
and the extra counterterm $\ratamp{1}{\gamma}{}{}$ 
reconstructs the finite
contribution 
stemming from
the part $\ntilde(\barq)$ of the numerator.
Since $\ratamp{1}{\gamma}{}{}$ terms arise only from UV divergences,
similarly as the usual UV counterterms
they originate only from UV-divergent one-particle irreducible (1PI) 
subdiagrams, where they take the form of polynomials of the external momenta and 
internal masses. Thus the insertions of 
$\ratamp{1}{\gamma}{}{}$ counterterms  into 
scattering amplitudes 
gives rise to rational functions of the 
kinematic invariants.%
\footnote{In the literature the $\ratamp{1}{\gamma}{}{}$ terms 
in \refeq{eq:ren1l4dim} are 
usually denoted as rational terms of type $R_2$,
and should not be confused with the so-called  
rational terms of type $R_1$.
See  \refse{se:oneloopratterms}
for more details.}

The goal of this paper is to extend the reconstruction of
$\ntilde$-contributions to two loops, such as to enable two-loop calculations
based on numerical tools that build the numerator of Feynman integrals in
four dimensions.
In our analysis we will focus 
on two-loop $\ntilde$-contributions of UV
origin assuming that IR divergences
are either absent, like in off-shell scattering amplitudes, or are subtracted in a way that 
does not generate rational terms.
A systematic analysis of $\ntilde$-contributions of IR origin 
is deferred to future work.

The reconstruction of $\ntilde$-contributions of UV origin will be 
carried out at the level of UV-renormalised
two-loop amplitudes. 
In renormalisable theories, the UV renormalisation 
can be implemented through a
recursive procedure that is known as the 
\textR-operation~\cite{Bogoliubov:1957gp,hepp1966,Zimmermann1969,Kennedy_ROp}
and amounts to the insertion of local subtraction terms into multi-loop
diagrams and their subdiagrams.
For the amplitude of a two-loop diagram 
$\Gamma$, the \textR-operation has the form
\bea 
\label{eq:twoloopRformula}
{\textbf{R}}\, \ampbar{2}{\Gamma}{}{}  
&=&  
\ampbar{2}{\Gamma}{}{} + \sum  \limits_{\gamma}  \deltaZ{1}{\gamma}{}{} 
\cdot \ampbar{1}{\Gamma/\gamma}{}{} + \deltaZ{2}{\Gamma}{}{}\,,
\eea
where $\ampbar{2}{\Gamma}{}{}$ is the 
unrenormalised two-loop amplitude in $\dendim$ dimensions,
and the remaining terms on the rhs correspond to a two-step 
subtraction.
In the first step, the
subdivergence of the various
one-loop subdiagrams $\gamma$ are subtracted 
by inserting the counterterms $\deltaZ{1}{\gamma}{}{}$
into  their complementary one-loop
diagrams $\Gamma/\gamma$, which 
are derived from $\Gamma$ by 
shrinking $\gamma$ to a vertex.
In the second step,
the remaining local two-loop divergence of $\Gamma$
is subtracted by the local counterterm $\deltaZ{2}{\Gamma}{}{}$.
The identity \refeq{eq:twoloopRformula}
is applicable also when $\Gamma$ 
is a set of two-loop diagrams.
In this case the bookkeeping of $\gamma$ and $\Gamma/\gamma$,
which can be single diagrams or sets of diagrams, 
follows naturally from the case of a single two-loop diagram by 
using $\bfR$ as a linear operation.
In fact, the 
\textR-operation
is typically applied 
at the level of 
full 1PI vertex functions
$\Gamma$ and $\gamma$.

As we will demonstrate,
the following
generalisation of the \textR-operation makes it possible 
to construct 
renormalised two-loop amplitudes 
using loop integrands with four-dimensional numerators 
and rational counterterms
for the reconstruction of $\ntilde$-contributions,
\bea 
\label{eq:masterformula} 
{\textbf{R}}\, \ampbar{2}{\Gamma}{}{}   
&=&  \amp{2}{\Gamma}{}{} + 
\sum  \limits_{\gamma} \lb \deltaZ{1}{\gamma}{}{} +\deltaZtilde{1}{\gamma}{}{} + \ratamp{1}{\gamma}{}{} \rb \cdot \amp{1}{\Gamma/\gamma}{}{}
+ \lb \deltaZ{2}{\Gamma}{}{} + \ratamp{2}{\Gamma}{}{} \rb\,.
\eea
Here the two-loop amplitude $\amp{2}{\Gamma}{}{}$ 
and its one-loop parts $\amp{1}{\Gamma/\gamma}{}{}$
are computed with four-dimensional 
numerators.
The $\msbar$ counterterms $\deltaZ{2}{\Gamma}{}{}$ 
and $\deltaZ{1}{\gamma}{}{}$ 
are related to the ones
in \refeq{eq:twoloopRformula} via 
trivial projection to four dimensions.
Quadratically
divergent one-loop
subdiagrams require additional counterterms
$\deltaZtilde{1}{\gamma}{}{}$, which subtract
extra poles of the form $\tildeqidx{}{2}/\eps$, with $\tilde q =
\barq-q$, that appear as a consequence of the different 
dimensionality of the loop momenta in the two-loop numerator and
denominator.
The one-loop UV counterterms are accompanied by 
related $\ratamp{1}{\gamma}{}{}$ counterterms,
which reconstruct the
$\ntilde$-contributions stemming from subdivergences.
Similarly, the two-loop UV counterterms are 
supplemented by $\ratamp{2}{\gamma}{}{}$ counterterms for 
the reconstruction of the remaining 
$\ntilde$-contributions, which originate from the local two-loop
divergences remaining after the subtraction of all subdivergences.

As we will 
show, 
the $\ratamp{2}{\Gamma}{}{}$ contributions
arise only from superficially divergent 1PI two-loop diagrams 
and can be reduced to a finite set
of process-independent local counterterms.
Using a tadpole decomposition
technique~\cite{Misiak:1994zw,beta_den_comp}, which is 
well known 
from the computation
of renormalisation constants and renormalisation group functions, we will derive a general
formula for the calculations of the $\delta \calR_2$ counterterms in any
renormalisable theory.
Finally, as a first application, we present the full set of two-loop rational
counterterms for QED in the $R_{\xi}$-gauge.

We note that the connection established in this paper between two-loop
amplitudes with loop numerators in $D$ and four dimensions bears some similarity
to the relations presented in~\cite{Page:2015zca} between two-loop QCD
vertex functions in dimensional regularisation and in the
four-dimensional regularisation/renormalisation (FDR)
approach~\cite{Pittau:2012zd}.
However, these two studies are based on very different 
regularisation and renormalisation procedures.
In the FDR approach loop integrals are entirely kept
in four dimensions, and the divergences are cancelled by means of a set of
subtraction rules.
In contrast, our approach is based on loop integrals in $D$ dimensions, where
only the numerator is restricted to four dimensions, and the contributions
stemming from its $(D-4)$-dimensional parts are reconstructed in a way that corresponds
exactly to $\msbar$-renormalised amplitudes in dimensional regularisation.
Moreover, we point out that the properties of UV rational terms established in
this paper are proven in a fully general way.

The paper is organised as follows. In \refse{se:notation}
we introduce our notation and 
conventions.  
In \refse{sec:oneloop} we review rational terms at one loop, and we
introduce the tadpole decomposition method
of~\cite{Misiak:1994zw,beta_den_comp},
which will be used to calculate rational counterterms and 
to discuss their general properties.
In \refse{se:ddimoneloop} we consider one-loop diagrams with
$\dendim$-dimensional external loop momenta and the related 
$\delta\tilde Z_{1}$ counterterms.
The master formula \refeq{eq:masterformula} 
for the reconstruction of rational terms 
is derived in \refse{se:irredtwoloop}, where  
we also present a general formula for 
the calculation of the required
$\delta \calR_2$ counterterms.
Explicit results for such counterterms in 
QED can be found in \refse{sec:qedres},
and the $\msbar$ counterterms for QED in the 
$R_\xi$ gauge are listed in \refapp{sec:UVct}.

\section{Notation and conventions}
\label{se:notation}
In this section we introduce our conventions for the treatment of 
dimensionally regularised scattering amplitudes 
and for 
their decomposition into 
irreducible loop subdiagrams
and tree subdiagrams.

\subsection{Notation for $\dendim$-dimensional quantities}

For the regularisation of UV divergences 
in this paper we use the 't~Hooft--Veltman scheme~\cite{tHooft:1972tcz}, 
where external states are four-dimensional, while 
loop momenta as well as the metric tensors and Dirac
matrices inside the loops live in 
\bea
\dendim=4-2\eps
\eea
dimensions.
For the analysis of rational terms we use an 
additional parameter $\numdim$, which
denotes the dimensionality of loop numerators 
and can take the values
\bea
\label{eq:ddimnotI}
\numdim &=&\begin{cases}
\dendim & \mbox{for calculations in $\dendim$ dimensions}\,, \\
4 & \mbox{for calculations with four-dimensional loop numerator\,.}
\end{cases}
\eea
Amplitudes with loop numerator in four dimensions and loop denominators in $\dendim$
dimensions will be referred to as $\numdim=4$ dimensional amplitudes.

In $\numdim=\dendim$ dimensions, all relevant ingredients of 
loop numerators will be decomposed into four-dimensional parts and 
($\dendim-4$)-dimensional remnants.
Contractions of Lorentz vectors in $\dendim$ dimensions are decomposed as
\bea
\label{eq:ddimnotB}
A_{\denbar\mu}B^{\denbar \mu} = A_{\mu}B^{\mu} + A_{\tilde \mu}B^{\tilde
\mu}\,,
\eea
where the indices $\bar \mu$ 
range 
over all components of the
$\dendim$-dimensional vectors,
the indices $\mu$ are restricted to four dimensions,
and the indices $\tilde \mu$ 
are associated with
the $(\dendim-4)$-dimensional remnant.
In general, to distinguish $\dendim$ and $(\dendim-4)$-dimensional quantities from their
four-dimensional counterparts  we use symbols carrying 
a bar and a tilde, respectively.
For the $\dendim$-dimensional loop momentum we write
\bea
\label{eq:ddimnotC}
\barq &=& q+\tilde q\,,
\eea
where
\bea
\label{eq:ddimnotD}
q^\mu = \barqidx{}{\mu},\qquad
\tildeqidx{}{\tilde \mu} = \barqidx{}{\tilde\mu}\,,
\eea
and
\bea
\label{eq:ddimnotD2}
\barqidx{}{2}= q^2+\tildeqidx{}{2}\,.
\eea
For the integration measure in loop-momentum space we use the
shorthand
\bea
\int\!\rd\barq & = & \mu^{2\eps} \int \f{\rd^{^D}\! \bar
q}{(2\pi)^{^D}}\,,
\eea
where $\mu$ is the scale of dimensional regularisation and 
will be identified with the renormalisation scale.

Given that $q^{\tilde \mu}=\tildeqidx{}{\mu}=0$,  
for the Lorentz indices of $q$ and $\tilde q$ we 
often use a sloppy notation where we identify
$q^{\denbar \mu}\equiv q^\mu$ and $\tildeqidx{}{\denbar \mu} \equiv \tildeqidx{}{\tilde
\mu}$. Thus \refeq{eq:ddimnotC} will be typically written as
\bea
\label{eq:ddimnotE}
\barqidx{}{\denbar \mu} = q^\mu + \tildeqidx{}{\tilde\mu}\,.
\eea
This leads to contractions of objects that carry different kinds
of indices and have to be understood as follows,
\bea
\label{eq:ddimnotF}
A_{\denbar\mu}B^{\mu} &=& A_{\mu}B^{\mu}\,,
\qquad
A_{\denbar\mu}B^{\tilde \mu} = A_{\tilde \mu}B^{\tilde\mu}\,,
\qquad
A_{\mu}B^{\tilde \mu} = 0\,.
\eea
A similar notation is used also for the decomposition of Dirac matrices and the metric
tensor,
\bea
\label{eq:ddimnotG}
\denbar \gamma^\mu &=& \gamma^\mu+\tilde \gamma^{\tilde\mu}\,,
\nonumber\\
\denbar g^{\denbar \mu\denbar \nu} &=& g^{\mu\nu}+\tilde g^{\tilde\mu\tilde
\nu}\,.
\eea
Metric tensors with indices of different type should be understood as 
\bea
\label{eq:ddimnotH}
\bar g^{\mu \bar \nu} &=& g^{\mu \bar\nu} \,=\,  g^{\mu \nu}\,.
\eea

\subsection{Reducible and irreducible loop amplitudes}
\label{se:irredloops}

\newcommand{\diaheightB}{25mm}
\newcommand{\diaheightBir}{31mm}

Our analysis of rational terms of UV origin will be carried out at the level 
of UV-renormalised amplitudes.
Before renormalisation, the amplitude of a one-loop diagram $\gamma$ has the
general form
\bea
\label{eq:irredloopA}
\fullamp_{1,\gamma} &=&
\vcenter{\hbox{
\includegraphics[height=\diaheightB]{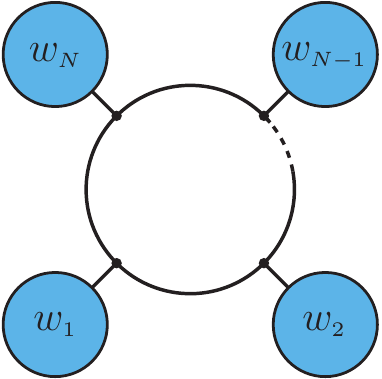}
}}
\,=\,
\ampbar{1}{\gamma}{\sigma_1\ldots \sigma_N}{}\,
\prod_{i=1}^{N} \left[w_{i}\right]_{\sigma_i}\,,
\eea
where $\ampbar{1}{\gamma}{}{}$ corresponds to the amplitude of the 1PI amputated 
one-loop subdiagram of $\gamma$, which is connected to the external lines 
through the factorised subtrees $w_{i}$,
depicted as blue bubbles.
We denote as subtree a tree subdiagram that connects an internal vertex
to a set of external lines.
Since external subtrees are free from UV singularities, 
only the 1PI subdiagram needs to be renormalised, \ie
\bea
\label{eq:irredloopC}
\bfR \, \fullamp_{1,\gamma}&=& 
\left(\bfR \, \ampbar{1}{\gamma}{\sigma_1\ldots \sigma_N}{}\right)\,
\prod_{i=1}^{N} \left[w_{i}\right]_{\sigma_i}\,.
\eea
Two-loop diagrams can be classified into two types depending on whether the 
topology that results from the amputation of all external subtrees is
irreducible or still reducible. 
The amplitude of a two-loop diagram $\Gamma$ of the first type has the form
\bea
\label{eq:irredloopE}
\fullamp_{2,\Gamma} &=&
\vcenter{\hbox{
\includegraphics[height=\diaheightBir]{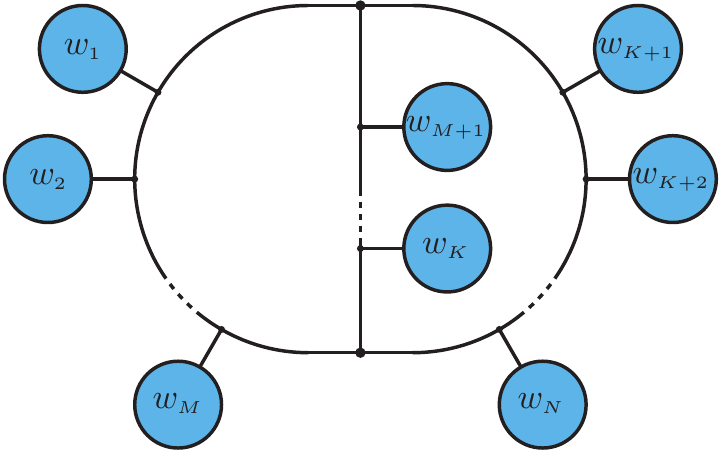}
}}
\,=\,
\ampbar{2}{\Gamma}{\sigma_1 \cdots \sigma_{\sss{N}}}{}\,
\prod_{i=1}^{N} \left[w_{i}\right]_{\sigma_i}\,,
\eea
where $\ampbar{2}{\Gamma}{}{}$  corresponds to
the amplitude of the 1PI amputated two-loop diagram
that is left after factorisation of 
all external subtrees $w_i$.
Similarly as in the one-loop case, the \textR-operation acts only on the 1PI
part,
\bea
\label{eq:irredloopF}
\bfR \, \fullamp_{2,\Gamma}
&=&
\left(\bfR \, \ampbar{2}{\Gamma}{\sigma_1 \cdots \sigma_{\sss{N}}}{}\right)\,
\prod_{i=1}^{N} \left[w_{i}\right]_{\sigma_i}\,.
\eea
The general form of the amplitude of a two-loop diagram $\Gamma_{\mathrm{red}}$ of the second type is
\def\gammared{\Gamma_{\mathrm{red}}}
\bea
\label{eq:irredloopG}
\fullamp_{2,\gammared} \!\! &=& \!\!\!\!
\vcenter{\hbox{
\includegraphics[height=\diaheightB]{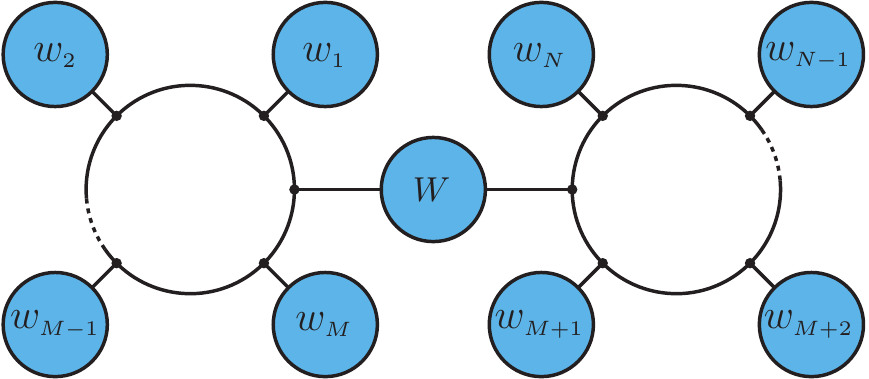}
}}
\!\!=\,
\ampbar{1}{\gamma_1}{\alpha_1\sigma_1 \cdots \sigma_{\sss{M}}}{}\,
W_{\alpha_{1}\alpha_{2}}\,
\ampbar{1}{\gamma_2}{\alpha_2\sigma_{M+1} \cdots \sigma_{\sss{N}}}{}\,
\prod_{i=1}^{N} 
\left[w_{i}\right]_{\sigma_i}\,.
\nonumber\\
\eea
Here, the factorisation of all external subtrees $w_i$ leads to 
two separate 1PI amputated one-loop amplitudes, 
$\ampbar{1}{\gamma_{1}}{}{}$ and $\ampbar{1}{\gamma_{2}}{}{}$,
that are connected to each other 
through a tree structure $W$.
Also in this case the \textR-operation acts only on the 
1PI building blocks,
\bea
\label{eq:irredloopD}
\bfR \, \fullamp_{2,\gammared}
&=& 
\left(\bfR \, \ampbar{1}{\gamma}{\alpha_1\sigma_1 \cdots \sigma_{\sss{M}}}{}\right)\,
W_{\alpha_{1}\alpha_{2}}
\left(\bfR \, \ampbar{1}{\gamma}{\alpha_2\sigma_{M+1} \cdots \sigma_{\sss{N}}}{}\right)\,
\prod_{i=1}^{N} \left[w_{i}\right]_{\sigma_i}\,.
\eea
In this paper we will consider $\msbar$ renormalised amplitudes in the
't~Hooft--Veltman
scheme, where all tree structures $w_i$ and $W$ 
in \refeq{eq:irredloopA}--\refeq{eq:irredloopD} 
are in four dimensions.
Thus the external momenta and external indices of the 1PI amplitudes
$\ampbar{1}{\gamma}{}{}$ and $\ampbar{2}\Gamma{}{}{}$
are handled as four-dimensional quantities.
Since they are free from $(\dendim-4)$-dimensional parts,
in the 't~Hooft--Veltman scheme  tree structures do not generate any rational
term.\footnote{As far as rational 
terms of UV origin are concerned,
the $(\dendim-4)$-parts of tree structures, which appear in
alternative schemes such as conventional dimensional regularisation, are 
irrelevant since they factorise 
with respect to the 
UV-renormalised 1PI subdiagrams.}
Thus rational terms can be determined 
at the level of 
1PI subdiagrams 
and directly extended to full amplitudes 
through \refeq{eq:irredloopC} and
\refeq{eq:irredloopF}--\refeq{eq:irredloopD}.
%

%%%%%%%%%%%%%%%%%%%%%%%%%%%%%%%%%%%%%%%%%%%%%%%%%%%%%%%
\section{Rational terms at one loop and the tadpole method} \label{sec:oneloop}
%%%%%%%%%%%%%%%%%%%%%%%%%%%%%%%%%%%%%%%%%%%%%%%%%%%%%%%

This section deals with the structure of rational terms at one loop and 
their connection with UV poles.
In this context we introduce a general technique
that makes it possible to reduce rational terms of UV origin
to tadpole integrals.

\subsection{Rational parts of one-loop diagrams}
\label{se:oneloopratterms}

Let us consider the amplitude of a one-particle irreducible
one-loop diagram $\gamma$,
\bea 
\label{eq:rtoneloopA}
\ampbar{1}{\gamma}{}{} &=& 
\int\!\rd\barq_1\, \f{{\barN}(\bar{q}_1)}{\Dbar{0}(\barq_1)\cdots
\Dbar{N-1}(\barq_1)}\,,
\eea
with denominators
\bea
\label{eq:rtoneloopB}
\Dbar{j}(\barq_1)&=& (\barq_1+p_j)^2-m_j^2\,,
\eea
where 
\bea
\label{eq:rtoneloopC}
p_j^\mu=\sum_{i=1}^j{k_i^\mu}\,,
\eea
and $k_1,\dots,k_N$ are the $N$ external momenta flowing into the loop.
Momentum conservation implies 
$\sum_{i=1}^N k_i=0$.
Colour structures and all Lorentz
or Dirac indices associated with the amputated external legs that enter the
loop are implicitly understood.  Such indices as well as all 
external momenta are treated as four-dimensional quantities
as discussed in \refse{se:irredloops}.

In $\numdim=\dendim$ dimensions, the numerator $\bar \calN(\barq_1)$ can be split into
\bea
\label{eq:rtoneloopD}
\bar \calN(\barq_1)&=& \calN(q_1) + \tilde \calN(\barq_1)\,,
\eea
where 
\bea
\label{eq:rtoneloopE}
\calN(q_1) &=& 
\bar \calN(\barq_1)\,\Big|_{
\bar g\to g,\, 
\bar \gamma\to \gamma\,, 
\barq_1 \to q_1
}
\eea
is the four-dimensional part,
obtained by projecting the metric tensor, Dirac matrices
and the loop momentum to four dimensions.
By construction, the remnant $\ntilde(\barq_1)$ vanishes in $\numdim=4$ dimensions. More
precisely,
\bea
\label{eq:rtoneloopF}
\tilde \calN(\barq_1) &=& \ord(\eps, \tilde q_1)
\eea
in $\numdim=4-2\eps$ dimensions. Thus we will refer to $\tilde \calN$
as the $(\dendim-4)$-dimensional part of the numerator.

At the level of the one-loop amplitude the splitting \refeq{eq:rtoneloopD}
results into 
\bea 
\label{eq:rtoneloopG}
\ampbar{1}{\gamma}{}{} &=& \amp{1}{\gamma}{}{} + \ratamp{1}{\gamma}{}{}\,,
\eea
where 
\bea 
\label{eq:rtoneloopH}
\calA_{1} &=& 
\int\!\rd\barq_1\, \f{\calN(q_1)}{\Dbar{0}(\barq_1)\cdots
\Dbar{N-1}(\barq_1)}\,
\eea
can be computed with tools that handle the numerator in
$\numdim=4$ 
dimensions while 
retaining the full $\dendim$-dependence of the 
loop momentum in the denominator.
The remnant part,
\bea 
\label{eq:rtoneloopI}
\ratamp{1}{\gamma}{}{} &=& 
\int\!\rd\barq_1\, \f{\tilde{\calN}(\barq_1)}{\Dbar{0}(\barq_1)\cdots
\Dbar{N-1}(\barq_1)}\,,
\eea
will be referred to as $\ntilde$-contribution.
Here the only relevant terms are the $\ord(\eps^0)$ 
contributions that originate from the interplay of the $(\dendim-4)$-dimensional 
part of the numerator with $1/\eps$ poles. 
At one loop such $\ntilde$-contributions originate only from 
poles of UV type~\cite{Bredenstein:2008zb}, 
and similarly as for UV poles
they arise only from
UV divergent 1PI
functions, where they take the form of simple polynomials in the external
momenta and internal masses.  
For this reason, $\ntilde$-contributions can be
reconstructed through a finite set of process-independent
counterterms~\cite{Ossola:2008xq,Draggiotis:2009yb,Garzelli:2009is,Pittau:2011qp}.
Their insertion into tree amplitudes gives rise to rational functions of the 
kinematic invariants.

In the literature, the one-loop terms that arise from the
$(\dendim-4)$-dimensional part of the loop denominators
in \refeq{eq:rtoneloopH} and from  $\ntilde$
are denoted, respectively, rational terms of type $R_1$ and $R_2$.
The rational terms of type $R_1$ emerge from the reduction of tensor integrals to scalar integrals
and can be handled with numerical algorithms in four dimensions (see
e.g.\cite{delAguila:2004nf,Ossola:2006us}).  
However, they can not be
reduced to a finite set of counterterms.
In this paper we will focus on the
rational 
terms that originate from $\ntilde$
at one and two loops. Since 
we consider a single type of rational terms, for convenience we
will use the symbols $\delta\calR_L$, where $L=1,2,\dots$
indicates the loop order and not the kind of rational term.
We will refer to such contributions
as $\ntilde$ rational terms or
simply rational terms.

We note that the relation~\refeq{eq:rtoneloopG} may be regarded as 
a regularisation-scheme transformations that connects 
the amplitude $\ampbar{1}{\gamma}{}{}$ in the
't~Hooft--Veltman scheme to its counterpart
$\amp{1}{\gamma}{}{}$  in a pseudo-regularisation scheme
corresponding to the prescription~\refeq{eq:rtoneloopE}. 
However, we point out that the four-dimensional projection~\refeq{eq:rtoneloopE}
breaks gauge invariance and cannot be regarded as a consistent regularisation 
prescription.
Only the combination of the two terms on the rhs of~\refeq{eq:rtoneloopG}
should be regarded as a consistently regularised amplitude, and---by
construction---this combination is equivalent to the 't~Hooft--Veltman scheme.
We also note that the prescription~\refeq{eq:rtoneloopE} should not be confused with the 
four-dimensional helicity scheme (FDH)~\cite{BERN1992451,Bern:2002zk},
where the $(D-4)$-dimensional part of the loop momentum 
is retained throughout.\footnote{At one loop,
scattering amplitudes in the FDH scheme 
can be reconstructed in terms of loop integrals with four-dimensional numerators 
using the FDF approach~\cite{Fazio:2014xea}.}

\subsection{Tadpole decomposition}
\label{se:tadpoledec}

In this section we discuss a general method~\cite{Misiak:1994zw,beta_den_comp,Zoller:2014xoa} 
that makes it
possible to cast the UV divergent parts of loop integrals---which are at the
origin of rational terms---in the form of tadpole integrals.
This method is first introduced for one-loop integrals, 
while its application to two-loop integrals is discussed in \refse{se:proof}.

For the analysis of UV divergences it is convenient to 
express one-loop amplitudes 
in terms of tensor integrals,
\bea
\label{eq:oneloopstructD}
T_N^{\bar\mu_1\cdots \bar\mu_r}
&=&
\int\rd\barq_1\, \frac{\barqidx{1}{\bar\mu_1}\cdots \barqidx{1}{\bar\mu_r}}
{D_0(\barq_1)\cdots D_{N-1}(\barq_1)}\,.
\eea
In the case of the one-loop amplitude \refeq{eq:rtoneloopA}
we have
\be
\label{eq:oneloopstructD2}
\barA_{1,\gamma}\,=\,\sum\limits_{r=0}^{R}\barN_{\bar\mu_1\cdots \bar\mu_r} T_N^{\bar\mu_1\cdots
\bar\mu_r}\,,
\ee
where the coefficients $\barN_{\bar\mu_1\cdots \bar\mu_r}$ 
depend on the external momenta and helicities, and
are related to the loop numerator via
\bea
\label{eq:oneloopstructD3}
\bar \calN(\barq_1)
&=&
\sum_{r=0}^R  \bar\calN_{\bar\mu_1\cdots \bar\mu_r}\,
\barqidx{1}{\bar\mu_1}\cdots \barqidx{1}{\bar\mu_r}\,.
\eea
The loop integrals \refeq{eq:oneloopstructD} give rise to 
UV singularities if their integrands scale like
$q^X$ with $X\ge 0$ at $q\to \infty$.
The power $X$ is referred to as superficial degree of divergence
and can be determined via naive power counting in $q$.
For the tensor integrals \refeq{eq:oneloopstructD} it 
is given by 
\bea
X=4+r-2N\,,
\eea
and the tensor rank $r$ fulfils $r\le R\le N$ in renormalisable theories.

In order to isolate UV poles, it is 
convenient to separate the loop denominators into leading and 
subleading UV parts according to
\bea
\label{eq:dentadexpA2}
D_k(\barq_1) 
&=& 
\left(\barqidx{1}{2}-M^2\right) 
-
\Delta_k(\barq_1)\,,
\eea
with
\bea
\label{eq:dentadexpA3}
\Delta_k(\barq_1)&=& 
-p_k^2-2 \barq_1\cdot p_k+m_k^2-M^2\,,
\eea
where $M$ 
is an auxiliary mass scale.\footnote{Note that only the squared scale $M^2$ appears.}
The dominant UV contribution of 
$\ord(q_1^2)$ is captured by the term $(\barqidx{1}{2}-M^2)$,
which corresponds to the form of a massive tadpole propagator, 
while $\Delta_k(\barq_1)$ is a 
subleading contribution of $\ord(q_1^1)$.
Note that for one-loop amplitudes with four-dimensional 
external momenta the $(\dendim-4)$-dimensional part of the loop momentum 
does not contribute to \refeq {eq:dentadexpA3}, \ie
$\Delta_k(\barq_1)=\Delta_k(q_1)$. In contrast, the
external momenta of a one-loop subdiagram that is embedded in a 
two-loop diagram can depend on the second loop momentum $\barq_2$, giving rise to 
$\dendim$-dimensional terms of the form $-\barqidx{2}{2}\pm 2\barq_1\cdot \barq_2$
in  $\Delta_k(\barq_1)$.

Inverting the lhs and the rhs of \refeq{eq:dentadexpA2} and using partial fractioning 
leads to the tadpole decomposition formula~\cite{Misiak:1994zw,beta_den_comp}
\bea
\label{eq:dentadexpA}
\frac{1}{D_k(\barq_1)}&=& 
\frac{1}{\barqidx{1}{2}-M^2}
+
\frac
{\Delta_k(\barq_1)}
{\barqidx{1}{2}-M^2}
\frac{1}{D_k(\barq_1)}\,,
\eea
which separates a generic scalar propagator
into a leading tadpole contribution of order
$1/q_1^2$ 
and a subleading remnant 
consisting of the original propagator times 
an extra suppression factor of order\footnote{For propagators with 
$p_k=0$  the extra suppression factor is of order $1/q_1^2$.}
$1/q_1$.
The identity \refeq{eq:dentadexpA} holds exactly,
and its recursive application makes it possible to generate 
a systematic expansion of the propagators in the limit $1/q_1\to 0$. More
explicitly, applying 
\refeq{eq:dentadexpA}  $X+1$ times yields 
\bea
\label{eq:dentadexpB2}
\frac{1}{D_k(\barq_1)}&=& 
\sum_{\sigma=0}^X
\frac
{\left[\Delta_k(\barq_1)\right]^{\sigma}}
{\left(\barqidx{1}{2}-M^2\right)^{\sigma+1}}
+
\frac
{\left[\Delta_k(\barq_1)\right]^{X+1}}
{\left(\barqidx{1}{2}-M^2\right)^{X+1}}
\frac{1}{D_k(\barq_1)}\,,
\eea
where the sum on the rhs consists of pure tadpole terms of order 
$1/ q_1^{2}$,\dots, $1/ q_1^{X+2}$
and corresponds to the first $X+1$
terms of the Taylor expansion of 
\bea
\label{eq:dentadexpB}
\frac{1}{D_k(\barq_1)}
&=& 
\frac{1}{\barqidx{1}{2}-M^2}
\left[1+\frac{\Delta_k(\barq_1)}{\barqidx{1}{2}-M^2}
\right]^{-1}
\eea
in the expansion parameter $\Delta_k(\barq_1)/(\barqidx{1}{2}-M^2)$.
The exact remnant of such a truncated expansion,
\ie all missing
contributions of order  $1/ q_1^{X+3}$ and higher, 
is captured by the term involving the original propagator
on the rhs of~\refeq{eq:dentadexpB2}.

In order to render \refeq{eq:dentadexpB2} 
and similar decomposition formulas 
more compact, we introduce 
two operators that generate
%, respectively,
the truncated expansion in $\Delta_k(\barq_1)/(\barqidx{1}{2}-M^2)$
and its remnant, respectively.
Specifically, for the two terms on the rhs of 
\refeq{eq:dentadexpB2} we write\footnote{The superscript in $\bfS^{(1)}_X$ and $\bfF^{(1)}_X$ 
refers to the chain of $q_1$-dependent propagator denominators on which the operator acts. 
}
\bea
\label{eq:dentadexpC}
\bfS^{(1)}_X \frac{1}{D_k(\barq_1)}&=& 
\sum_{\sigma=0}^X
\frac
{\left[\Delta_k(\barq_1)\right]^{\sigma}}
{\left(\barqidx{1}{2}-M^2\right)^{\sigma+1}}\,,
\qquad
\bfF^{(1)}_X \frac{1}{D_k(\barq_1)}
\,=\,
\frac
{\left[\Delta_k(\barq_1)\right]^{X+1}}
{\left(\barqidx{i}{2}-M^2\right)^{X+1}}
\frac{1}{D_k(\barq_1)}\,.
\eea
More generally, at the level of the full one-loop integrand the above operators 
are defined as an exact decomposition,
\bea
\label{eq:chaintadexpA0}
\bfS^{(1)}_X+ \bfF^{(1)}_X &=& 1\,,
\eea
and they act only on  the $q_1$-dependent chain of loop denominators, \ie 
\bea
\label{eq:chaintadexpA}
\bfS^{(1)}_X\,
\frac{\bar\calN(\barq_1)}{D_0(\barq_1)\cdots D_{N-1}(\barq_1)}%\nonumber\\
&=& 
\bar\calN(\barq_1)\,
\bfS^{(1)}_X\,
\frac{1}{D_0(\barq_1)\cdots D_{N-1}(\barq_1)}
\,,
\eea
and similarly for $\bfF^{(1)}_X=1-\bfS^{(1)}_X$.
These two operators define 
a tadpole decomposition up to order $X$ of the entire chain of propagators, 
where $\bfS^{(1)}_X$ collects all pure tadpole terms with 
denominators $\left(\barqidx{1}{2}-M^2\right)^{N+\sigma}$ and $\sigma\in[0,X]$, while $\bfF^{(1)}_X$ corresponds to the remnant.
More precisely,
$\bfS^{(1)}_X$ amounts to a Taylor expansion 
of the full chain of propagators up to total order $X$
in the various $\Delta_k(\barq_1)/(\barqidx{1}{2}-M^2)$, \ie
\bea
\label{eq:chaintadexpA2}
\bfS^{(1)}_X 
\frac{1}{D_0(\barq_1)\cdots D_{N-1}(\barq_1)}&=& 
\sum_{\sigma=0}^{X}
\frac
{\Delta^{(\sigma)}(\barq_1)}
{\left(\barqidx{1}{2}-M^2\right)^{N+\sigma}}\,,
\eea
with
\bea
\label{eq:chaintadexpB}
{\Delta^{(\sigma)}(\barq_1)}
&=& 
\sum_{\sigma_0=0}^\sigma
\ldots
\sum_{\sigma_{N-1}=0}^\sigma
\prod_{k=0}^{N-1}
\left[\Delta_k(\barq_1)\right]^{\sigma_k}
\Bigg|_{\sigma_0+\dots+ \sigma_{N-1} = \sigma}\,.
\eea
Thus $\bfS^{(1)}_X$ turns the original integrals into a combination of
massive tadpole integrals
that include all terms from order $1/q_1^{2N}$ 
to order $1/q_1^{(2N+X)}$.
The numerators $\Delta^{(\sigma)}(\barq_1)$
of such tadpole integrals are 
polynomials of degree 
$\sigma$ in $\barq_1\cdot p_k$ and in the 
squared mass scales $\{p_k^2\}$, $\{m_k^2\}$ and $M^2$.
By construction, the remainder part associated with 
$\bfF^{(1)}_X$ involves only terms where 
the original degree of UV singularity is reduced by 
$X+1$ or more, \ie formally
\bea
\label{eq:chaintadexpB2}
\bfF^{(1)}_X \,\le \, \ord\left(\frac{1}{q^{X+1}}\right)\,.
\eea
Note also that the $\bfF^{(1)}_X$ remainder embodies all possible IR
poles of the original integral since 
the $\bfS^{(1)}_X$ operator converts the denominators of all massless propagators 
into massive ones.

In practice, the expansion \refeq{eq:chaintadexpA2}--\refeq{eq:chaintadexpB}
can be generated by applying the decomposition 
\refeq{eq:dentadexpA} in a recursive way 
until terms with denominators of the form 
\bea
\left(\barqidx{1}{2}-M^2\right)^{p}
D_{j_1}(\barq_1)\cdot\cdot\cdot D_{j_q}(\barq_1) 
\eea
with $p+q>N+X$ are encountered, and attributing 
such terms to $\bfF^{(1)}_X$.
Note that, according to the above definition of the 
tadpole expansion, the $\bfS^{(1)}_X$ operator
captures all terms up to relative order $1/q_1^X$
but retains also unnecessary terms of higher order in $1/q_1$.
This is due to the fact that terms of $\ord(q_1^1)$ 
and $\ord(q_1^0)$ in \refeq{eq:dentadexpA3} are treated on the same footing.
Possible optimisations based on power counting 
in $1/q_1$ and other tricks are briefly discussed in 
\refse{se:proof}.

For integrals with UV degree of divergence $X$, 
contributions that are suppressed by a relative factor $1/q^{X+1}$ 
do not contribute to the
divergence. Thus, using \refeq{eq:chaintadexpA}--\refeq{eq:chaintadexpA2}
we can express the pole part of the tensor integral \refeq{eq:oneloopstructD}
in terms of tadpole integrals with  one auxiliary mass scale $M$,
\bea
\label{eq:oneloopdivC}
\bfK\, T_N^{\bar\mu_1\cdots \bar\mu_r}
&=&
\bfK\, \bfS^{(1)}_X T_N^{\bar\mu_1\cdots \bar\mu_r}
\,=\, 
\sum_{\sigma=0}^{X}
\bfK
\int\rd\barq_1\, 
\frac
{\barqidx{1}{\bar\mu_1}\cdots\barqidx{1}{\bar \mu_r}\,
\Delta^{(\sigma)}(\barq_1)}
{\left(\barqidx{1}{2}-M^2\right)^{N+\sigma}}
\,.
\eea
Here and in the following $\bfK$ should be understood as 
a linear operator that isolates the 
%$\msbar$ 
pole part of an integral and discards the finite remnant. 
More precisely, let us consider the typical form of the
Laurent series that result  from $L$-loop integrals,
\bea
f_L(\eps)&=& 
S^{L\eps} \sum_{k=1}^L  \frac{\fmsbar{k}}{\eps^{k}} +\fmsbar{0} +
\ord(\eps)
\,=\,
\sum_{k=1}^L  \frac{\fms{k}}{\eps^{k}} +\fms{0} + \ord(\eps)\,,
\eea
where 
\bea
\label{eq:msbarnorm}
S^\eps &=& \left(4\pi\right)^\eps \Gamma(1+\eps)
%
%\left(4\pi\, e^{-\gamma_\mathrm{E}}\right)^\eps
%
\,=\, 1+ \eps\left[\ln(4\pi)-\gamma_{\mathrm{E}}\right]+\ldots
\eea
is the well-known  universal factor associated with each 
loop-momentum integration.
In the MS scheme the $\bfK$ operator is defined as 
\bea
\bfK\,f_L(\eps)
&\mseq&
\sum_{k=1}^L  \frac{\fms{k}}{\eps^{k}}\,,
\eea
while in the $\msbar$ scheme it should be understood as
\bea
\bfK\,f_L(\eps)
&\msbareq&
S^{L\eps} \sum_{k=1}^L  \frac{\fmsbar{k}}{\eps^{k}}\,.
\eea

Since the full tadpole decomposition \refeq{eq:chaintadexpA0} is independent
of $M$, and the truncated $\bfF_X^{(1)}$ part 
does not contribute to the divergence, 
the $M$-dependence of the tadpole integrals cancels on the
rhs of \refeq{eq:oneloopdivC}. Moreover, the general form of 
\refeq{eq:chaintadexpA2}--\refeq{eq:chaintadexpB}
implies that the pole residues  
are homogenous polynomials of
degree $X$ in the external momenta and internal masses.

The above tadpole decomposition can be
used also at two loops (and beyond). To this end,
as detailed in \refse{se:proof}, 
two-loop integrals are split into the three chains of propagators that depend on the
loop momenta $q_1$, $q_2$ and $q_3=-q_1-q_2$, and
two-loop divergencies are extracted by means of 
three separate tadpole decompositions 
with operators $\bfS_{X_i}^{(i)}=1-\bfF_{X_i}^{(i)}$
that act on the particular chain of $q_i$-dependent denominators,
for $i=1,2,3$,
and are otherwise
defined as in \refeq{eq:chaintadexpA0}--\refeq{eq:chaintadexpB2}.

\subsection{One-loop poles and rational parts in terms of tadpole integrals}
\label{se:onelooppoles}

In order to highlight the connection between UV poles and rational
$\ntilde$-contributions, we
introduce an operator $\bar \bfK$ that extracts the full contribution of UV
poles at the level of one-loop amplitudes in $\numdim=\dendim$ dimensions.
For the generic one-loop amplitude \refeq{eq:rtoneloopA}, using the
tensor decomposition \refeq{eq:oneloopstructD2}, we define the 
$\bar \bfK$ operator as 
\bea
\label{eq:oneloopdivD}  
\bar \bfK\, \ampbar{1}{\gamma}{}{} 
&=&
\bar \bfK\,
\sum_{r=0}^R  \bar\calN_{\bar\mu_1\cdots \bar\mu_r}
T_N^{\bar\mu_1\cdots \bar\mu_r}\,=\,
\sum_{r=0}^R  \bar\calN_{\bar\mu_1\cdots \bar\mu_r}\,
\bfK\,T_N^{\bar\mu_1\cdots \bar\mu_r}\,,
\eea
and we split it 
into two pieces,
\bea
\label{eq:oneloopdivD1}    
\bar \bfK\, \ampbar{1}{\gamma}{}{} 
&=&
\bfK\, \ampbar{1}{\gamma}{}{}+
\tilde \bfK\, \amp{1}{\gamma}{}{}\,,
\eea
which result, respectively, from the interplay of the UV poles
$\bfK\,T_N^{\bar\mu_1\cdots \bar\mu_r}$
with the four-dimensional and $(\dendim-4)$-dimensional parts of 
$\bar\calN_{\bar\mu_1\cdots \bar\mu_r}$.
The former yields the UV singularity
\bea
\label{eq:oneloopdivE}
\bfK\, \ampbar{1}{\gamma}{}{}
&=&
\sum_{r=0}^R  \calN_{\mu_1\cdots \mu_r}\,
\bfK\,
T_N^{\mu_1\cdots \mu_r}
\,=\,
-\deltaZ{1}{\gamma}{}{}\,,
\eea
where $\deltaZ{1}{\gamma}{}{}$ is the 
UV 
counterterm for the amplitude at hand,
while the $(\dendim-4)$-dimensional part of the numerator 
gives rise to the $\ntilde$-contribution%
\footnote{On the lhs of \refeq{eq:oneloopdivF} we write 
$\amp{1}{\gamma}{}{}$ without bar
since, a posteriori, $\tilde \bfK\, \amp{1}{\gamma}{}{}$ 
can be reconstructed from the 
$\ratamp{1}{\gamma}{}{}$ counterterm using only 
four-dimensional ingredients. 
However, it should be clear that, a priori,
$\tilde \bfK\, \amp{1}{\gamma}{}{}$ 
depends on the $(\dendim-4)$-dimensional 
part of the loop numerator.
}
\bea
\label{eq:oneloopdivF}
\bfKtilde\, \amp{1}{\gamma}{}{}
&=&
\sum_{r=0}^R  
\left[\bar \calN_{\bar\mu_1\cdots \bar\mu_r}-
\calN_{\mu_1\cdots \mu_r}
\right]\,
\bfK\, 
T_N^{\bar\mu_1\cdots \bar\mu_r}
\,=\,
\ratamp{1}{\gamma}{}{}
\,.
\eea
Note that the difference within square brackets 
can be regarded as the combination of two kinds of 
$(\dendim-4)$-dimensional terms:
a contribution \mbox{$\bar \calN_{\bar\mu_1\cdots \bar\mu_r}-
\bar \calN_{\mu_1\cdots \mu_r}$} that 
originates from the $(\dendim-4)$-dimensional 
components of the loop momentum in the 
tensor integrals \refeq{eq:oneloopstructD}, and a second contribution
\mbox{$\bar \calN_{\mu_1\cdots \mu_r} - \calN_{\mu_1\cdots \mu_r}$}
that corresponds to the remaining $(\dendim-4)$-dimensional part of the 
loop numerator.

In renormalisable theories UV singularities at one loop arise only from 
diagrams with 
$N\le 4$
loop propagators. 
Thus, UV poles and $\ntilde$-contributions can be
derived once and for all at the level of
the relevant 1PI vertex functions
and encoded in a finite set of
$\deltaZ{1}{\gamma}{}{}$ and $\ratamp{1}{\gamma}{}{}$ 
counterterms.
The identities \refeqs{eq:oneloopdivE}{eq:oneloopdivF}
can be regarded as the master formulas for the 
derivation of such counterterms.
To this end, the poles of tensor integrals can be computed in terms of
tadpole integrals using \refeq{eq:oneloopdivC}. As discussed above, 
the residues of such poles are $M$-independent polynomials of the external momenta $\{p_k\}$ 
and internal masses $\{m_k\}$. 
As a consequence, at the level of 1PI vertex functions,
$\deltaZ{1}{\gamma}{}{}$ and $\ratamp{1}{\gamma}{}{}$ 
are local counterterms. More precisely, they 
take the form of homogeneous polynomials of
degree $X$ in the external momenta $\{p_k\}$ and internal masses $\{m_k\}$, 
while their insertion at the level of full
scattering amplitudes yields rational functions of the kinematic
invariants\cite{Ossola:2008xq,Draggiotis:2009yb,Garzelli:2009is,Pittau:2011qp}.

In \refse{se:irredtwoloop}, using a similar strategy based on tadpole
decompositions and power counting, we demonstrate that also
two-loop $\ntilde$-contributions of UV origin can be reconstructed by means
of a finite set of local
counterterms.

%%%%%%%%%%%%%%%%%%%%%%%%%%%%%%%%%%%%%%%%%%%%%%%%%%%%%%%%%%%%%%%%%%%%%
\section{One-loop diagrams with $\dendim$-dimensional external momenta}
\label{se:ddimoneloop}
%%%%%%%%%%%%%%%%%%%%%%%%%%%%%%%%%%%%%%%%%%%%%%%%%%%%%%%%%%%%%%%%%%%%%

\begin{figure}%[t]
\begin{center}
\includegraphics[height=0.15\textheight]{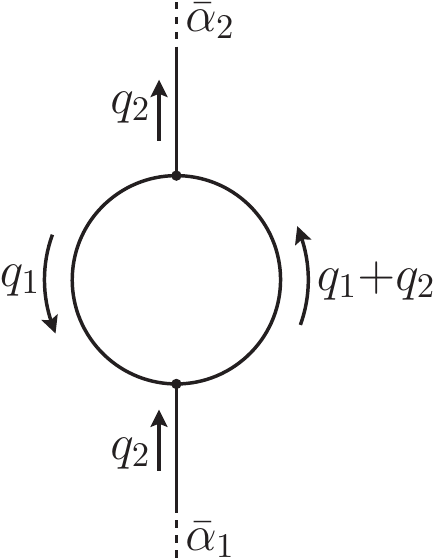}
\hspace{20mm}
\includegraphics[height=0.15\textheight]{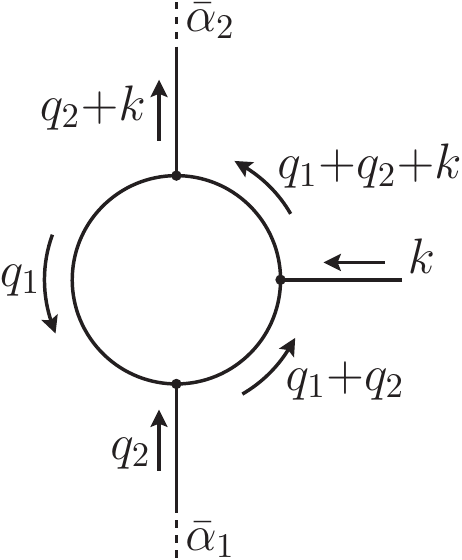}
\end{center}
\caption{One-loop subtopologies that 
can
give rise to non-logarithmic UV
divergences.}
\label{fig:oneloopsubdiag}
\end{figure}

As a preparation for the discussion of $\ntilde$-contributions at two loops, in this
section we extend the analysis of one-loop UV poles and $\ntilde$-terms
to the case of one-loop subdiagrams of two-loop diagrams.
Specifically, as depicted in \reffi{fig:oneloopsubdiag}, we consider one-loop subdiagrams with internal loop
momentum $\barq_1$ and two external lines that 
depend on the loop momentum $\barq_2$
and are going to be embedded in a two-loop diagram.
In the following $\barq_2$ is kept
fixed, and we investigate the role of its $(\dendim-4)$-dimensional
part $\tilde q_2$. In particular, we show that 
non-logarithmic UV subdivergences can give rise to 
non-trivial contributions of the form $\tildeqidx{2}{2}/\eps$.

\subsection{One-loop subdiagram in $\numdim=\dendim$ dimensions}
\label{se:ddimsubdiag}

Let us consider one-loop subdiagrams of the type shown in 
\reffi{fig:oneloopsubdiag}. The corresponding loop numerators have the form
\bea
\label{eq:ddimsubdiagA}
\bar \calN^{\baralpham}(\barq_1,\barq_2)
&=&
\sum_{r=0}^R  \bar\calN_{\bar\mu_1\cdots \bar\mu_r}^{\baralpham}(\barq_2)\,
\barqidx{1}{\bar\mu_1}\cdots \barqidx{1}{\bar\mu_r}\,,
\eea
where $\barq_1$ is the loop momentum  of the
subdiagram at hand, $\barq_2$ is the external loop momentum, and the
multi-index $\baralpham = (\bar\alpha_1, \bar\alpha_2)$ combines the two
Lorentz/Dirac indices associated with the two $\barq_2$-dependent external
lines. 

For what concerns UV poles and $\ntilde$-contributions, as long as the dimensionality
of $\barq_2$ is the same in the loop numerator and
denominator, the analysis of \refse{se:onelooppoles} is applicable to the
case at hand via naive extension of the external degrees of freedom from 
four to $\dendim$ dimensions.
More explicitly, the formulas~\refeq{eq:oneloopdivD}--\refeq{eq:oneloopdivF}
take the form,
\bea
\label{eq:ddimsubdiagB}  
\bar \bfK\, \ampbar{1}{\gamma}{\baralpham}{\barq_2} 
&=&
\bfK\, \ampbar{1}{\gamma}{\baralpham}{\barq_2} +
\tilde \bfK\, \amp{1}{\gamma}{\baralpham}{\barq_2}
\,=\,
\sum_{r=0}^R  \bar\calN_{\bar\mu_1\cdots \bar\mu_r}^{\baralpham}(\barq_2)\,
\bfK\,
T_N^{\bar\mu_1\cdots \bar\mu_r}(\barq_2)\,,
\eea
with the UV divergent part
\bea
\label{eq:ddimsubdiagC}
\bfK\, \ampbar{1}{\gamma}{\baralpham}{\barq_2} 
&=&
\sum_{r=0}^R  \calN_{\mu_1\cdots \mu_r}^{\baralpham}(\barq_2)\,
\bfK\,
T_N^{\mu_1\cdots \mu_r}(\barq_2)
\,=\,
-\deltaZ{1}{\gamma}{\baralpham}{\barq_2}\,,
\eea
and the $\ntilde$-part
\bea
\label{eq:ddimsubdiagD}
\tilde \bfK\, \amp{1}{\gamma}{\baralpham}{\barq_2} 
&=&
\sum_{r=0}^R  
\left[\bar \calN_{\bar\mu_1\cdots \bar\mu_r}^{\baralpham}(\barq_2)-
\calN_{\mu_1\cdots \mu_r}^{\baralpham}(\barq_2)
\right]\,
\bfK\, 
T_N^{\bar\mu_1\cdots \bar\mu_r}(\barq_2)
\,=\,
\ratamp{1}{\gamma}{\baralpham}{\barq_2}
\,.
\eea
The tensor integrals $T_N^{\bar\mu_1\cdots \bar\mu_r}(\barq_2)$ are defined
as in \refeq{eq:oneloopstructD}, and their $\barq_{2}$-dependence 
originates entirely from the loop denominators.
All quantities in \refeq{eq:ddimsubdiagB}--\refeq{eq:ddimsubdiagD}, 
including the counterterms 
$\deltaZ{1}{\gamma}{\baralpham}{\barq_2}$
and
$\ratamp{1}{\gamma}{\baralpham}{\barq_2}$,
are polynomials of degree $X$ in $\barq_2$,
and are related to the corresponding quantities in~\refeq{eq:oneloopdivD}--\refeq{eq:oneloopdivF}
through the replacements
$q_2\to \barq_2$ and $\alpham\to \bar \alpham$.
For later convenience we also rewrite \refeq{eq:oneloopdivF} as
\bea
\label{eq:ddimsubdiagD2}
\tilde \bfK\, \amp{1}{\gamma}{\alpham}{q_2} 
&=&
\sum_{r=0}^R  
\left[\bar \calN_{\bar\mu_1\cdots \bar\mu_r}^{\alpham}(q_2)-
\calN_{\mu_1\cdots \mu_r}^{\alpham}(q_2)
\right]\,
\bfK\, 
T_N^{\bar\mu_1\cdots \bar\mu_r}(q_2)
\,=\,
\ratamp{1}{\gamma}{\alpham}{q_2}
\,,
\eea
where the 
dependence on $q_2$ and $\alpha$
is made explicit.  Since \refeq{eq:oneloopdivF}
and \refeq{eq:ddimsubdiagD} are free from UV poles, as long as $q_2$ is
not integrated they differ only by terms of order $(\dendim-4)$.  More precisely,
\bea
\label{eq:ddimsubdiagD3}
\tilde \bfK\, \ampbar{1}{\gamma}{\baralpham}{\barq_2} 
&=&
\tilde \bfK\, \amp{1}{\gamma}{\alpham}{q_2} + \ord(\eps,\tilde q_2)
\,.
\eea

As an example
of a one-loop diagram with $\dendim$-dimensional
external momentum,
let us consider the massless photon selfenergy
in QED,
\bea
\ampbar{1}{\gamma}{\bar\alpha_1\bar\alpha_2}{\barq_2} & = & 
-\ri e^2\int\!\rd\barq_1 
\frac{\Tr\big[\bar\gamma^{\bar \alpha_1}\slashed{\barq}_1 \bar\gamma^{\bar\alpha_2} (\slashed{\barq}_1 + \slashed{\barq}_2)\big]}{\barqidx{1}{2} \, ( \barq_1+\bar
q_2)^2}\,.
\eea
In this case, the quadratic UV divergence generates 
quadratic polynomials of the external momentum $\barq_2$, 
\bea
\bar\bfK\,\ampbar{1}{\gamma}{\bar\alpha_1\bar\alpha_2}{\barq_2} 
&
\mseq
& 
\f{\ri\alpha}{4\pi}\left[
-\f{4}{3\, \eps}\lb \barqidx{2}{2} g^{\denbar\alpha_1\denbar\alpha_2} 
- \barqidx{2}{\denbar\alpha_1} \barqidx{2}{\denbar\alpha_2} \rb
+
\f{2}{3}\, \barqidx{2}{2} \, g^{\denbar\alpha_1\denbar\alpha_2}
\right]\,,
\eea
where the two terms between square brackets correspond, respectively, to the 
UV
pole~\refeq{eq:ddimsubdiagC} 
and the rational $\ntilde$-contribution~\refeq{eq:ddimsubdiagD}.
Note that for the examples discussed in this section we adopt the
MS scheme, while the final results in \refse{sec:qedres}
are presented in the $\msbar$ scheme.

\subsection{One-loop subdiagram in $\numdim=4$ dimensions}
\label{se:4dimsubdiag}

In order to identify the $\ntilde$-contributions that originate from 
one-loop subdiagrams, in the following we compare the
$\dendim$-dimensional numerator
\refeq{eq:ddimsubdiagA} to its four-dimensional variant,
\bea
\label{eq:4dimsubdiagA}
\calN^{\alpham}(q_1,q_2)
&=&
\sum_{r=0}^R  \calN_{\mu_1\cdots \mu_r}^{\alpham}(q_2)\,
q_1^{\mu_1}\cdots q_1^{\mu_r}\,,
\eea
where 
all parts of the numerator, including
the multi-index $\alpham$ 
and the external loop momentum $q_2$, 
are
projected to four dimensions.
At the amplitude level, 
in analogy with \refeq{eq:ddimsubdiagB},   
the interplay of the numerator \refeq{eq:4dimsubdiagA} with
the UV poles that arise from the $\barq_1$-integration 
results into
\bea
\label{eq:4dimsubdiagB}
\bar \bfK\,\amp{1}{\gamma}{\alpham}{q_2}
&=&
\bfK\,\amp{1}{\gamma}{\alpham}{q_2}
\,=\,
\sum_{r=0}^R  \calN^{\alpham}_{\mu_1\cdots
\mu_r}(q_2)
\,\bfK\,T_N^{\mu_1\cdots \mu_r}(\barq_2)\,,
\eea
where the tensor integrals $T_N^{\mu_1\cdots \mu_r}(\barq_2)$ depend
on $\barq_2$ since the external loop momentum is kept in $\dendim$
dimensions in the loop denominator.
The full pole contribution~\refeq{eq:4dimsubdiagB} can be split into 
two parts, 
\bea
\label{eq:4dimsubdiagB2}
\bfK \amp{1}{\gamma}{\alpham}{q_2}
&=& 
- \deltaZ{1}{\gamma}{\alpham}{q_2}
- \deltaZtilde{1}{\gamma}{\alpham}{\tilde q_2}\,,
\eea
where the first part reads
\bea
\label{eq:4dimsubdiagC}
- \deltaZ{1}{\gamma}{\alpham}{q_2}
&=&
\sum_{r=0}^R  \calN^{\alpham}_{\mu_1\cdots
\mu_r}( q_2)
\,\bfK\,T_N^{\mu_1\cdots \mu_r}(q_2)\,
\eea
and corresponds to the
standard 
UV 
counterterm~\refeq{eq:ddimsubdiagC}
with $(\bar\alpha,\barq_2)$ replaced by  $(\alpha, q_2)$
throughout.
The remnant part originates from the $(\dendim-4)$-dimensional part of 
$\barq_2$ in the denominator of the one-loop subdiagram
and reads
\bea
\label{eq:4dimsubdiagE}
- \deltaZtilde{1}{\gamma}{\alpham}{\tilde q_2}\,
&=&
\sum_{r=0}^R  \calN^{\alpham}_{\mu_1\cdots
\mu_r}( q_2)
\,
\Delta K_N^{\mu_1\cdots \mu_r}(\tilde q_2)\,,
\eea
with 
\bea
\label{eq:ztildeformA}
\Delta K_N^{\mu_1\cdots \mu_r}(\tilde q_2)
&=&
\bfK\,T_N^{\mu_1\cdots \mu_r}(\barq_2)-
\bfK\, T_N^{\mu_1\cdots \mu_r}(q_2)\,.
\eea
In renormalisable theories, where the maximum degree of divergence of one-loop
integrals is $X=2$, the tensor-integral poles in \refeq{eq:ztildeformA}
are at most quadratic in $\barq_2$. Their general form is
\bea
\label{eq:ztildeformB}
\bfK\,T_N^{\mu_1\cdots \mu_r}(\barq_2)
&\mseq&
\frac{1}{\eps}\left[A^{\mu_1\cdots \mu_r}+
B_{\bar\nu_1}^{\mu_1\cdots \mu_r}\,\barqidx{2}{\bar\nu_1}+
C_{\bar\nu_1\bar\nu_2}^{\mu_1\cdots \mu_r}\,\barqidx{2}{\bar\nu_1}
\barqidx{2}{\bar\nu_2}\right],
\eea
where the tensors $A$, $B$ and $C$ consist of combinations of 
the other four-dimensional external momenta $p_k$ and metric tensors, 
which carry four-dimensional indices $\mu_i$ or $\dendim$-dimensional
indices $\bar \nu_j$.
In \refeq{eq:ztildeformA} the $\barq_2$-independent contribution associated
with the tensor $A$ cancels, and
for integrals with degree of divergence $X\le 1$,
where the tensor $C$ associated with the quadratic $\barq_2$ terms
vanishes,
we have 
\bea
\label{eq:ztildeformC}
\Delta K_N^{\mu_1\cdots \mu_r}(\tilde q_2) \bigg|_{X\le 1}
&\mseq&
\frac{1}{\eps}
B_{\bar\nu_1}^{\mu_1\cdots \mu_r}\,\tildeqidx{2}{\bar\nu_1}
\,=\,0\,.
\eea
This cancellation is due to the fact that the tensor $B$ 
carries a single $\dendim$-dimensional index, which 
can lead only to terms of the form 
$p_k\cdot \tilde q_2=0$ or $g^{\mu_i}_{\tilde \nu_1} \tildeqidx{2}{\bar
\nu_1}=\tildeqidx{2}{\mu_i}= 0$.
Thus \refeq{eq:ztildeformA} is non-vanishing only for 
quadratically divergent integrals. In this case
\bea
\label{eq:ztildeformD}
\Delta K_N^{\mu_1\cdots \mu_r}(\tilde q_2) \bigg|_{X = 2}
&\mseq&
\frac{1}{\eps}
C_{\bar\nu_1\bar\nu_2}^{\mu_1\cdots \mu_r}\,
\left(\barqidx{2}{\bar\nu_1}\barqidx{2}{\bar\nu_2}-
q_2^{\nu_1}q_2^{\nu_2}\right)
\nonumber\\
&=&
\frac{1}{\eps}
C_{00}^{\mu_1\cdots \mu_r}g_{\bar \nu_1\bar \nu_2}\,
\left(\barqidx{2}{\bar\nu_1}\barqidx{2}{\bar\nu_2}-
q_2^{\nu_1}q_2^{\nu_2}\right)
\,=\,
C_{00}^{\mu_1\cdots \mu_r}
\frac{\tildeqidx{2}{2}}{\eps}\,,
\eea
where we have split the tensor $C$ into a part $C^{\mu_1\dots\mu_r}_{00} g_{\bar \nu_1\bar \nu_2}$
and a remaining part that does not contribute to \refeq{eq:ztildeformD} 
since one or both $\bar \nu_i$ indices are either carried by a four-dimensional external
momentum or by a $g_{\bar \nu_i}^{\mu_j}$ tensor.
Based on \refeq{eq:ztildeformC}--\refeq{eq:ztildeformD} we conclude that in
renormalisable theories 
the extra counterterms  \refeq{eq:4dimsubdiagE}
are required only for
quadratically divergent selfenergies, and their general form is
\bea
\label{eq:4dimsubdiagE8}
\deltaZtilde{1}{\gamma}{\alpham}{\tilde q_2}
&\mseq&
v^\alpham \frac{\tildeqidx{2}{2}}{\eps}\,,
\eea
where $v^\alpha$ is independent of $q_2$.
Such extra counterterms 
should be regarded as an extension of 
the usual UV counterterms for the case of one-loop integrals with 
numerators in $\numdim=4$ and denominators in $\dendim=4-2\eps$
dimensions.
Note that upon integration over $\barq_2$
the terms of order $\tildeqidx{2}{2}/\eps$ in~\refeq{eq:4dimsubdiagE8}
result into two-loop contributions of order $\eps^0$. 

As an example of the UV poles of a one-loop diagram 
in $\numdim=4$ dimensions,
let us consider again
the massless photon selfenergy in QED,
\bea
\amp{1}{\gamma}{\alpha_1\alpha_2}{q_2} & = & 
-\ri e^2 
\int\!
\rd\barq_1
\frac{\Tr\big[\gamma^{\alpha_1}\slashed q_1 \gamma^{\alpha_2} (\slashed q_1 + \slashed q_2)\big]}{\barqidx{1}{2} \, ( \barq_1+\bar
q_2)^2}\,,
\eea
where
\bea
\bfK\,\amp{1}{\gamma}{\alpha_1\alpha_2}{q_2} 
&\mseq&
\f{\ri \alpha}{4\pi}\left[
-\f{4}{3\, \eps}\lb q_2^2 g^{\alpha_1\alpha_2} - q_2^{\alpha_1} q_2^{\alpha_2} \rb
-
\f{2}{3} \f{\tildeqidx{2}{2}}{\eps}\, g^{\alpha_1\alpha_2}
\right]\,.
\eea
Here the two terms between square brackets correspond, respectively, to the 
standard 
UV 
counterterm~\refeq{eq:4dimsubdiagC}
and the $\ord(\tilde q_2/\eps)$ remnant~\refeq{eq:4dimsubdiagE}.

\subsection{Relating renormalised one-loop subdiagrams in $\numdim=\dendim$ and 
$\numdim=4$}
\label{se:ddimoneloopsubt}

In this section we extend the identity~\refeq{eq:ren1l4dim} 
to one-loop amplitudes with $D$-dimensional external
momenta. As a starting point
we consider the amplitude of a renormalised 
subdiagram in $\numdim=\dendim$ dimensions,
\bea
\label{eq:rensubdiag}
\bfR\,\ampbar{1}{\gamma}{\baralpham}{\barq_2}
&=&
(1-\bfK)\,\ampbar{1}{\gamma}{\baralpham}{\barq_2}
\,=\,
\ampbar{1}{\gamma}{\baralpham}{\barq_2}
+
\deltaZ{1}{\gamma}{\baralpham}{\barq_2}\,,
\eea
and we relate it to corresponding quantities in 
$\numdim=4$ dimensions by means of rational terms.
To this end, using \refeq{eq:ddimsubdiagB}
as 
an auxiliary
subtraction term in $\numdim=\dendim$ 
we define the subtracted amplitude 
\bea
\label{eq:subdivX1}
\ampbar{1}{\gamma}{\baralpham}{\barq_2}
-
\bar\bfK\,\ampbar{1}{\gamma}{\baralpham}{\barq_2}
&=&
\sum_{r=0}^R  
\bar\calN^{\baralpham}_{\bar\mu_1\cdots \bar\mu_r}(\barq_2)\,
\Big[T_N^{\bar\mu_1\cdots \bar\mu_r}(\barq_2)
-\bfK\,T_N^{\bar\mu_1\cdots \bar\mu_r}(\barq_2)
\Big]\,.
\eea
Similarly, using \refeq{eq:4dimsubdiagB} as a subtraction term in
$\numdim=4$ we define
\bea
\label{eq:subdivX2}
\amp{1}{\gamma}{\alpham}{q_2}
-
\bfK\,\amp{1}{\gamma}{\alpham}{q_2}
&=&
\sum_{r=0}^R  
\calN^{\alpham}_{\mu_1\cdots \mu_r}(q_2)\,
\Big[T_N^{\mu_1\cdots \mu_r}(\barq_2)
-\bfK\,T_N^{\mu_1\cdots \mu_r}(\barq_2)
\Big]\,.
\eea
By construction, in both cases the subtraction terms cancel the full 
pole contribution at the level of tensor integrals.
Thus the terms between square 
brackets in \refeq{eq:subdivX1} and \refeq{eq:subdivX2}
are free from $1/\eps$ poles and differ only by terms of $\ord(\eps)$.
As a consequence, also the whole subtracted amplitudes differ only by terms of 
order $(\eps)$. More explicitly,
\bea
\label{eq:subdivK}
\ampbar{1}{\gamma}{\baralpham}{\barq_2}
-
\bar\bfK\,\ampbar{1}{\gamma}{\baralpham}{\barq_2}
&=&
\amp{1}{\gamma}{\alpham}{q_2}
-
\bfK\,\amp{1}{\gamma}{\alpham}{q_2} +\ord(\eps,\tilde q_2)\,.
\eea
This identity can be turned into a relation between 
renormalised amplitudes 
by splitting $\bar\bfK$ into $\bfK+\bfKtilde$ as in  \refeq{eq:ddimsubdiagB}, 
using \refeq{eq:ddimsubdiagD3} for the $\bfKtilde$ part, and
shifting the latter to the lhs. In this way one arrives at
\bea
\label{eq:subdivP}
(1-\bfK)\,\ampbar{1}{\gamma}{\baralpham}{\barq_2}
&=&
(1-\bfK+\bfKtilde)\,\amp{1}{\gamma}{\alpham}{q_2}
+\ord(\eps,\tilde q_2)
\,.
\eea
Finally, expressing the $\bfK$ and $\tilde \bfK$ terms through the
corresponding UV and rational
counterterms introduced in \refeq{eq:ddimsubdiagC}, 
\refeq{eq:ddimsubdiagD2} and \refeq{eq:4dimsubdiagB2}
leads to
\bea
\label{eq:subdivN}
\bfR\,\ampbar{1}{\gamma}{\baralpham}{\barq_2}
&=& 
\amp{1}{\gamma}{\alpham}{q_2}
+
\deltaZ{1}{\gamma}{\alpham}{q_2}
+
\deltaZtilde{1}{\gamma}{\alpham}{\tilde q_2}
+
\ratamp{1}{\gamma}{\alpham}{q_2}
+\ord(\eps,\tilde q_2 )\,.
\eea
This identity relates the 
UV-renormalised
amplitude in $\dendim$
dimensions, on the lhs, to the corresponding amplitude 
with four-dimensional numerator plus 
three counterterms:
the 
usual
UV
counterterm
$\deltaZ{1}{\gamma}{\alpham}{}$ 
with $\alpha$ and $q_2$
in four dimensions, 
its $\ord(\tildeqidx{2}{2}/\eps)$ extension
$\deltaZtilde{1}{\gamma\,}{\alpham}{}$,
defined in~\refeq{eq:4dimsubdiagE}--\refeq{eq:4dimsubdiagE8}, 
and the rational term $\ratamp{1}{\gamma\,}{\alpham}{}$, which
compensates for the 
missing $\ntilde$-part of the loop numerator.
At two loops, the identity~\refeq{eq:subdivN} 
will play a key role for the extraction of 
UV poles and $\ntilde$-contributions that arise from 
divergent one-loop subdiagrams
(see \refse{se:irredtwoloop}).

\begin{figure}[t]
\begin{center}
\bea
\bfR\;\left[ \;
\vcenter{\hbox{\includegraphics[height=\diaheight]{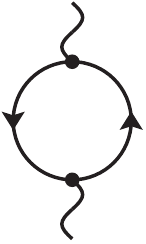}}} \;
\right]_{\numdim\,=\,\dendim}
\hspace{-2mm}
&=& 
\left[ \;
\vcenter{\hbox{\includegraphics[height=\diaheight]{QEDgpropIL}}}
\;\;+\;\;
\vcenter{\hbox{\includegraphics[height=\diaheight]{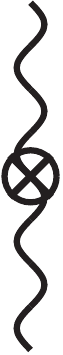}}}
\;\deltaZ{1}{\gamma}{}{}\,
\right]_{\numdim\,=\,\dendim}
\nonumber\\[4mm]
&=& 
\left[\; \vcenter{\hbox{\includegraphics[height=\diaheight]{QEDgpropIL}}}
\;\;+\;\;
\vcenter{\hbox{\includegraphics[height=\diaheight]{QEDgpropCT}}}
\;\Big(\deltaZ{1}{\gamma}{}{}
+
\deltaZtilde{1}{\gamma}{}{} 
+
\ratamp{1}{\gamma}{}{}
\Big)
\right]_{\numdim\,=\,4}  
\nonumber
\eea
\end{center}
\caption{Graphical representation of~\refeq{eq:rensubdiag} 
for a renormalised QED selfenergy in
$\numdim=\dendim$ numerator dimensions and the relation~\refeq{eq:subdivN} 
to its counterpart in $\numdim=4$ dimensions.
The indices $\alpha=(\alpha_1,\alpha_2)$,
the external loop momentum $q_2$ and terms of $\ord(\eps)$ 
are implicitly understood.
} 
\label{fig:rensubdiag}
\end{figure}

The renormalised amplitude \refeq{eq:rensubdiag} in $\numdim=\dendim$ and the
identity~\refeq{eq:subdivN} are illustrated in \reffi{fig:rensubdiag} for
the case of a QED selfenergy.

\section{Rational terms at two loops}
\label{se:irredtwoloop}

In this section we derive a general formula for the reconstruction of the
$\ntilde$-contributions of two-loop amplitudes in any 
renormalisable theory. 
We also present an explicit recipe for the calculation
of the relevant process-independent two-loop 
counterterms in terms of tadpole integrals.

\begin{figure}[t]
\begin{center}
\includegraphics[width=0.3\textwidth]{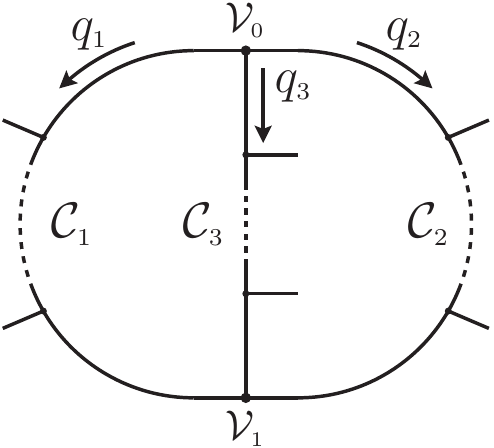}
\end{center}
\caption{A generic irreducible two-loop diagram consists of 
two vertices, $\calV_0$, $\calV_1$, that connect 
three chains, $\calC_1$, $\calC_2$, $\calC_3$, which 
contain, respectively, all propagators that depend on the loop momenta $q_1$, $q_2$,
$q_3=-q_1-q_2$, as well as all vertices that connect the propagators 
depending on the same loop momentum.
Note that in general the loop momenta 
that flow out of the
$\calV_0$ vertex should be $q_i+p_{i0}$, but for 
triple vertices one can set $p_{i0}=0$ as in the above graph.
In contrast, if $\calV_0$ is a quartic vertex,
momentum conservation requires 
non-vanishing $p_{i0}$ momenta with $\sum_i p_{i0}$
equal to the external momentum that flows into 
$\calV_0$.
} 
\label{fig:twoloop_irred}
\end{figure}

\subsection{Notation for two-loop diagrams and subdiagrams}
\label{se:twoloopnot}

Irreducible two-loop diagrams involve propagators that depend on the loop momenta
$q_1$, $q_2$ and $q_3=-q_1-q_2$. Their generic structure is 
illustrated in \reffi{fig:twoloop_irred} and consists of three
chains, $\calC_{1},\calC_{2},\calC_{3}$, that are connected to each
other by two vertices, $\calV_0,\calV_1$.
Each chain $\calC_{i}$ includes a certain number $N_i$ of
propagators
that depend on the loop
momentum $q_i$ and the $N_i-1$ vertices that connect them to each other and
to external lines.
The two-loop integral associated with a generic two-loop 
diagram $\Gamma$ has the form\footnote{This 
two-particle irreducible amplitude corresponds to 
$\ampbar{2}{\Gamma}{\sigma_1\dots \sigma_N}{}$ in \refeq{eq:irredloopF},
but here and in the following the external indices $\sigma_1\dots \sigma_N$ 
are kept implicit. 
}

\bea
\label{eq:twoloopnotA1}
\ampbar{2}{\Gamma}{}{} &=&
\int\rd\barq_1
\int \rd\barq_2
\int \rd\barq_3
\,\delta(\barq_1+\barq_2+\barq_3)\,
\frac{
\bar\Gamma^{\bar\alpha_1\bar\alpha_2\bar\alpha_3}(\barq_1,\barq_2,\barq_3)\,
\bar\calN^{(1)}_{\bar\alpha_1}(\barq_1)\,
\bar\calN^{(2)}_{\bar\alpha_2}(\barq_2)\,
\bar\calN^{(3)}_{\bar\alpha_3}(\barq_3)}
{\calD{1}\,\calD{2}\,\calD{3}}\,,
\nonumber\\
\eea 
where each chain $\calC_i$ contributes through the corresponding set of loop
denominators,
\bea
\label{eq:twoloopnotB}
\calD{i}&=&
D^{(i)}_0(\barq_i)\cdots
D^{(i)}_{N_i-1}(\barq_i)\,,
\qquad\mbox{with}\quad
D^{(i)}_a(\barq_i) \,=\, \left(\barq_i+p_{ia}\right)^2-m_{ia}^2\,,
\eea
and a loop numerator part $\bar\calN^{(i)}_{\bar\alpha_i}(\barq_i)$.
The latter carries a multi-index
$\bar \alpha_i\equiv (\bar \alpha_{i1},\bar \alpha_{i2})$
that connects the two ends of the 
chain to the tensor $\bar \Gamma^{\bar\alpha_1\bar \alpha_2\bar \alpha_3}$,
which embodies the two vertices, $\calV_0$ and $\calV_1$.
Integrating \refeq{eq:twoloopnotA1}
over $\barq_3$ yields 
\bea
\ampbar{2}{\Gamma}{}{} 
&=&
\int\rd\barq_1
\int \rd\barq_2\,
\frac{
\bar\calN(\barq_1,\barq_2)}
{\calD{1}\,\calD{2}\,\mathcal{D}^{(3)}(-\barq_1-\barq_2)}\,,
\eea 
where $\bar \calN(\barq_1,\barq_2)$ 
corresponds to the numerator of~\refeq{eq:twoloopnotA1}
at $\barq_3=-\barq_1-\barq_2$.
Similarly as in~\refeq{eq:rtoneloopD}--\refeq{eq:rtoneloopF} the two-loop
numerator can be split into
four- and $(\dendim-4)$-dimensional parts as
\bea
\label{eq:rttwoloopDD}
\bar \calN(\barq_1, \barq_2)&=& \calN(q_1, q_2) + \tilde \calN(\barq_1, \bar
q_2)\,,
\eea
where
\bea
\label{eq:rttwoloopEE}
\calN(q_1, q_2) &=& 
\bar \calN(\barq_1, \barq_2)\,\Big|_{
\bar g\to g,\, 
\bar \gamma\to \gamma,\,
\barq_1 \to q_1,\,
\barq_2 \to q_2
}
\eea
and
\bea
\label{eq:rttwoloopFF}
\tilde \calN(\barq_1,\barq_2) &=& \ord(\eps, \tilde q_1, \tilde q_2)\,.
\eea
The main goal of this paper is to derive a general formula for the
reconstruction of all relevant $\ntilde$-contributions of UV origin, \ie
all terms of order $\eps^{-1}$ and $\eps^0$ that originate form the
interplay the $(\dendim-4)$-dimensional 
part of the numerator \refeq{eq:rttwoloopFF} with single and double 
$1/\eps$ poles of UV type.

The analysis of $\ntilde$-contributions beyond one loop requires a careful treatment
of subdiagrams and their UV divergences.
At two loops, each diagram $\Gamma$ involves three subdiagrams, $\gamma_1$,
$\gamma_2$,  $\gamma_3$, where $\gamma_i$
results from $\Gamma$ by cutting the chain $\calC_i$.
More precisely, each partition $i | jk$ of $123$ corresponds to 
a subdiagram $\gamma_i$ that involves the chains $\calC_j$, $\calC_k$
and the vertices $\calV_0$, $\calV_1$. 
Its amplitude reads
\bea
\label{eq:subdiagnotA}
\ampbar{1}{\gamma_i}{\bar\alpha_i}{\barq_i} &=&
\int\rd\barq_j
\int \rd\barq_k
\,\delta(\barq_i+\barq_j+\barq_k)\,
\frac{
\bar\Gamma^{\bar\alpha_1\bar\alpha_2\bar\alpha_3}(\barq_1,\barq_2,\barq_3)\,
\bar\calN^{(j)}_{\bar\alpha_j}(\barq_j)\,
\bar\calN^{(k)}_{\bar\alpha_k}(\barq_k)}
{\calD{j}\,\calD{k}}\,,
\nonumber\\[2mm]
&=&
\int\rd\barq_j\,
\frac{
\bar\Gamma^{\bar\alpha_1\bar\alpha_2\bar\alpha_3}(\barq_1,\barq_2,\barq_3)\,
\bar\calN^{(j)}_{\bar\alpha_j}(\barq_j)\,
\bar\calN^{(k)}_{\bar\alpha_k}(\barq_k)}
{\calD{j}\,\calD{k}}\Bigg|_{\barq_k=-\barq_i-\barq_j}\,,
\eea
where $\barq_i$ plays the role of external momentum for the subdiagram 
$\gamma_i$.

For each subdiagram $\gamma_i$ of $\Gamma$ we define its 
complement $\Gamma/\gamma_i$ as the 
one-loop diagram that involves only the chain $\calC_i$
and results form $\Gamma$ by shrinking the chains
$\calC_j$, $\calC_k$ to a vertex.
Thus the full two-loop diagram $\Gamma$ can be expressed as
the insertion of the subdiagram $\gamma_i$ into its complement
$\Gamma/\gamma_i$.
For such insertions we use the notation
\bea
\label{eq:subdiagnotB}
\ampbar{2}{\Gamma}{}{}&=&
\ampbar{1}{\gamma_i}{}{}\cdot
\ampbar{1}{\Gamma/\gamma_i}{}{}
\,=\,
\int\rd\barq_i\,
\ampbar{1}{\gamma_i}{\bar\alpha_i}{\barq_i}\,
\frac{\bar\calN^{(i)}_{\bar\alpha_i}(\barq_i)}{\calD{i}}
\,,
\eea
where the dot product involves the integration over
the loop momentum $q_i$ and the summation over the multi-index $\alpha_i$,
as defined on the rhs.

\subsection{Power counting and structure of UV divergences}
\label{se:twolooppowcount}

Divergences of UV type can be easily identified through naive power
counting, \ie by 
counting the maximum power 
in the loop momenta $q_i$ at the integrand level.
For the analysis of two-loop divergences we count the powers of loop momenta originating from 
the loop chains $\calC_1$, $\calC_2$, $\calC_3$ 
and the connecting vertices $\calV_0$, $\calV_1$ as follows.
For a two-loop diagram $\Gamma$ we define 
$U_i(\Gamma)$ as the maximum power of the full chain 
$\calC_i$ in the corresponding 
loop momentum $q_i$
at $q_i\to\infty$. In QCD we have 
\bea
\label{eq:Xdef1}
U_i(\Gamma)&=&-\nprop{qq}{i}-2 \nprop{gg}{i}-2
\nprop{uu}{i}+\nvert{ggg}{i}+\nvert{uug}{i}\,, 
\eea
where the various terms on the rhs represent the numbers of 
propagators ($n_\prop$) and vertices ($n_\vert$) 
involving quarks ($q$), gluons ($g$) and ghosts ($u$)
along the chain $\calC_i$. The loop-momentum power of the 
vertices  $V_a$ that connect the three loop chains 
is denoted $Y_a(\Gamma)$.
In QED we have $Y_a(\Gamma)=0$, while in QCD
\bea
Y_a(\Gamma)&=&\begin{cases}
1 & \mbox{if $V_a$ is a trilinear gluon vertex}\,,\\
0 & \mbox{otherwise}\,.\\
\end{cases}
\eea
The loop momenta associated with ghost--gluon vertices can always be assigned to a unique
chain $\calC_i$, also in case of an intersection vertex $\calV_a$.
Thus $g u\bar u$ vertices should be accounted for through the corresponding counter
$\nvert{uug}{i}$ in $U_i(\Gamma)$ and not through $Y_a(\Gamma)$.

The simplest divergences of two-loop diagrams $\Gamma$ are the ones 
arising from their one-loop subdiagrams $\gamma_i$. They can be 
identified by means of the degree of subdivergence
\bea
\label{eq:Xijdef}
X(\gamma_i) &=& X_{jk}(\Gamma) 
\,=\, 4+ U_j(\Gamma) +U_k(\Gamma) +
\sum_{a=0}^1 Y_a(\Gamma)\,,
\eea
where $i|jk$ is a partition of 123.
When $X(\gamma_i)\ge 0$ the subdiagram $\gamma_i$
leads to a UV pole that arises from the region where
$q_j,q_k\to \infty$ with $q_i$ fixed.
Upon subtraction of all subdivergences, two-loop diagrams
can involve residual local divergences.
This kind of divergences originate from the region where \mbox{$q_1, q_2, q_3 \to \infty$}
simultaneously. They can be identified by means of the
global degree of divergence 
\be
\label{eq:Xdef}
X(\Gamma) \,=\, 8+\sum_{i=1}^3 U_i(\Gamma) +\sum_{a=0}^1 Y_a(\Gamma)\,,
\ee
which corresponds to the total loop-momentum power of 
the full two-loop diagram.
Diagrams with $X(\Gamma)\ge 0$ will be referred to 
as globally divergent diagrams. Such diagrams can 
involve both subdivergences and residual local divergences. 
Instead, diagrams with $X(\Gamma)< 0$ are free from local divergences and
in renormalisable theories they involve at most one 
subdivergence, \ie 
\be
\label{eq:maxonesubdivA}
X(\Gamma)\,<\,0 \qquad\Rightarrow \quad
X(\gamma_l)\,<\,0\quad\mbox{for at least two subdiagrams $\gamma_l$}\,.
\ee
This well-known property\footnote{The fact that $X(\Gamma)<  0$ implies at most one
subdivergence can be demonstrated 
by combining \refeq{eq:Xijdef} and \refeq{eq:Xdef} 
such as to obtain 
the following relation between 
the global degree of divergence $X(\Gamma)$
and the sum of the degrees of divergence of two arbitrary subdiagrams,
\bea
\label{eq:maxonesubdivB}
X(\gamma_j)+X(\gamma_k) &=& 
X(\Gamma)+ \tilde X_i(\Gamma)\,  
\eea
with
\bea
\label{eq:maxonesubdivC}
\tilde X_i(\Gamma)&=& U_i(\Gamma)+Y_0(\Gamma)+Y_1(\Gamma)\,.  
\eea
The latter quantity describes the total $q_i$-power of the 
$N_i$ propagators of the chain $\calC_i$ 
together with all (internal and external) $N_i+1$ vertices to which they are 
connected.
In renormalisable theories, each of the $N_i$ combinations 
of a propagator with a neighbouring 
vertex contributes with power $-1$ or less, while the 
remaining extra vertex contributes with power $+1$ or less.
Therefore
\bea
\label{eq:maxonesubdivD}
\tilde X_i(\Gamma)&\le&  -N_i+1 \,\le\, 0\quad\mbox{for}\quad N_i\ge 1.
\eea
Combining \refeq{eq:maxonesubdivD} with \refeq{eq:maxonesubdivB} yields 
the following lower bound for the global degree of divergence,
\bea
X(\Gamma) 
&\ge&
X(\gamma_j)+X(\gamma_k)\,,
\eea
which implies, consistently with \refeq{eq:maxonesubdivA},
that two or more subdivergences can occur only if
$X(\Gamma)>0$.
}
will be exploited in \refse{se:proof}
in order to 
demonstrate that 
\mbox{$\ntilde$-contributions}
at  two loops 
can be reconstructed by means of 
universal local counterterms.

\subsection{Structure of UV poles at two loops}
\label{se:polestructure}

In general, two-loop amplitudes involve one-loop subdivergences and
additional local divergences.
The renormalisation of these two kinds of divergences
can be schematically written in the form
\bea
\label{eq:twolooprenA}
\bfR\,\ampbar{2}{\Gamma}{}{}
&=&
\ampbar{2}{\Gamma}{}{}-
\bfKsub\, \ampbar{2}{\Gamma}{}{}-
\bfKloc\, \ampbar{2}{\Gamma}{}{}\,.
\eea
Here the $\bfKsub$ operator extracts the 
divergences that result from the 
$\msbar$ poles of the three subdiagrams,
\bea
\label{eq:twolooprenB}
\bfKsub\, \ampbar{2}{\Gamma}{}{}&=&
\sum_{i=1}^3
(\bfK\, \ampbar{1}{\gamma_i}{}{})\cdot \ampbar{1}{\Gamma/\gamma_i}{}{}
\,=\,
-\sum_{i=1}^3
\int\rd\barq_i\,
\deltaZ{1}{\gamma_i}{\bar\alpha_i}{\barq_i}\,
\frac{\bar\calN^{(i)}_{\bar\alpha_i}(\barq_i)}{\calD{i}}
\,.
\eea
The one-loop $\msbar$ counterterms $\delta Z_{1,\gamma_i}(\barq_i)$ 
are local, \ie they are polynomials in $\barq_i$,
but their insertion into two-loop diagrams
gives rise to non-local terms.
After subtraction of all one-loop subdivergences, two-loop 
diagrams with $X(\Gamma)\ge 0$ still involve 
local divergences. 
The corresponding $\msbar$ poles are 
extracted through the 
operator $\bfKloc$,
which is defined as
\bea
\label{eq:twolooprenC}
\bfKloc\, \ampbar{2}{\Gamma}{}{}&=&
\bfK\,(1-\bfKsub)\, \ampbar{2}{\Gamma}{}{}
\,=\,
-\deltaZ{2}{\Gamma}{}{}\,
\,,
\eea
where $\delta Z_{2,\Gamma}$ represents the 
two-loop  
UV
counterterm
for the diagram at hand.

Note that the renormalisation operator $\bfR$
and the associated 
operators $\bfKsub$ and $\bfKloc$ 
should be understood as linear operators.
In particular, 
when a two-loop diagram $\Gamma$ is split into a sum of contributions $\Gamma_\sigma$,
the $\bfKsub$ operator fulfils 
\bea
\label{eq:twolooprenD}
\bfKsub\, \left(\sum_\sigma \ampbar{2}{\Gamma_\sigma}{}{}\right)&=&
\sum_\sigma \bfKsub\, \ampbar{2}{\Gamma_\sigma}{}{}
\,,
\eea
with 
\bea
\label{eq:twolooprenE}
\bfKsub\, \ampbar{2}{\Gamma_\sigma}{}{}&=&
\sum_{i=1}^3
(\bfK\, \ampbar{1}{\gamma_{\sigma i}}{}{})\cdot
\ampbar{1}{\Gamma_\sigma/\gamma_{\sigma i}}{}{}
\,.
\eea
Here $\gamma_{\sigma i}$ denotes the $i^{\mathrm{th}}$ subdiagram associated 
with the contribution $\Gamma_\sigma$, 
and $\bfK\, \ampbar{1}{\gamma_{\sigma i}}{}{}$
is the corresponding UV pole.

\begin{figure}[t]
\begin{center}
\bea
\bfR\,
\left[\;
\vcenter{\hbox{\includegraphics[width=\diawidth]{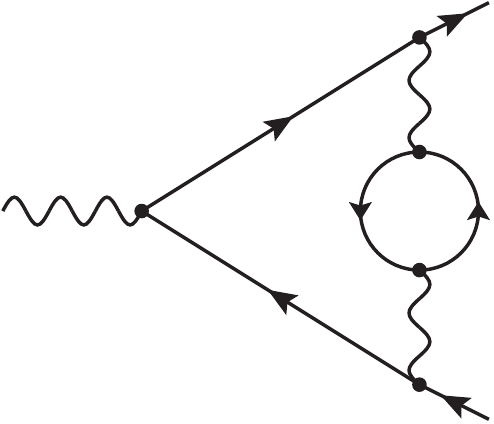}}}\;\;\right]_{{\numdim\,=\,\dendim}}
\hspace{-3mm}
&=& \!\! 
\left[\;
\vcenter{\hbox{\includegraphics[width=\diawidth]{QEDvtxIILoop}}}  
\;\;+\;
\vcenter{\hbox{\includegraphics[width=\diawidth]{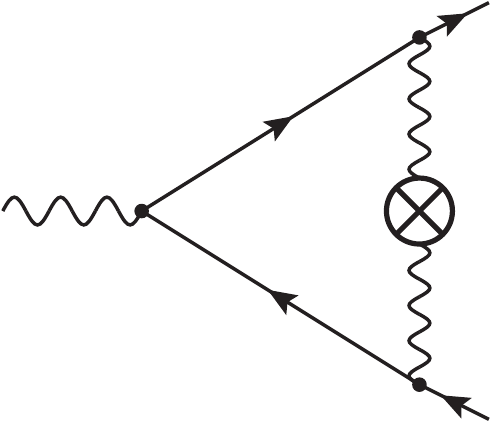}}}{
\deltaZ{1}{\gamma_i}{}{} } 
\;+\;
\vcenter{\hbox{\includegraphics[width=\diawidth]{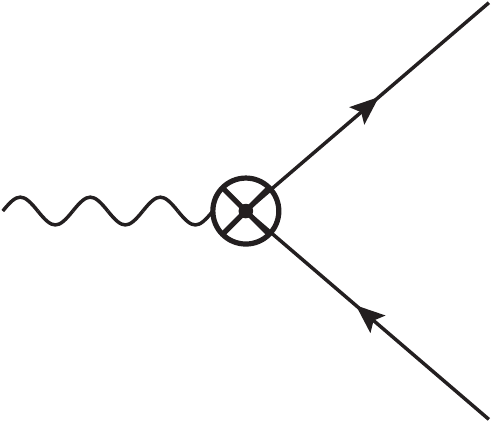}}} \hspace{-5mm}\deltaZ{2}{\Gamma}{}{} 
\;\right]_{\numdim\,=\,\dendim}
\nonumber
\eea
\end{center}
\caption{Graphical representation of the renormalisation formula \refeq{eq:twolooprenA} 
for a two-loop QED diagram with a single subdivergence.} 
\label{fig:twoloopren}
\end{figure}

The structure of the renormalisation formula 
\refeq{eq:twolooprenA} 
is illustrated in  \reffi{fig:twoloopren}
for the case of a two-loop QED diagram with a single subdivergence.

\subsection{Structure of rational parts at two loops}
\label{se:ratstructure}

The main goal of this paper is to derive a general formula that relates  
renormalised two-loop amplitudes in $\numdim=\dendim$ dimensions to corresponding 
amplitudes 
in $\numdim=4$ dimensions by means of rational counterterms.
This formula will be derived in \refse{se:proof}, and in the following we
anticipate its general structure,  which reads
\bea
\label{eq:twoloopratstA}
\bfR\,\ampbar{2}{\Gamma}{}{} 
&=&
\amp{2}{\Gamma}{}{}+
\left(\bfKtildesub-\bfKsub\right) \amp{2}{\Gamma}{}{}
+\left(\bfKtildeloc-\bfKloc\right) \amp{2}{\Gamma}{}{}\,.
\eea
Here the subtraction of subdivergences and local
two-loop divergences is implemented through the operators $\bfKsub$ and $\bfKloc$ in
a similar way as in \refeq{eq:twolooprenA}, but such operators are
supplemented by the $\bfKtildesub$ and $\bfKtildeloc$ operators, which
reconstruct the $\ntilde$-contributions that originate form the 
respective types of divergences.
Similarly as for $\bfKsub$ and $\bfKloc$, also
$\bfKtildesub$ and $\bfKtildeloc$ should be understood as linear operators in the sense of 
\refeq{eq:twolooprenD}--\refeq{eq:twolooprenE}.

According to our analysis in \refse{se:ddimoneloop}, 
the $\bfKsub$ operator 
needs to be defined as
\bea
\label{eq:twoloopratstB}
\bfKsub\, \amp{2}{\Gamma}{}{}&=&
\sum_{i=1}^3
(\bfK\, \amp{1}{\gamma_i}{}{})\cdot \amp{1}{\Gamma/\gamma_i}{}{}
\,=\,
-\sum_{i=1}^3
\int\rd\barq_i\,
\left[
\deltaZ{1}{\gamma_i}{\alpha_i}{q_i}\,
+
\deltaZtilde{1}{\gamma_i}{\alpha_i}{\tilde q_i}\,
\right]
\frac{\calN^{(i)}_{\alpha_i}(q_i)}{\calD{i}}
\,,\nonumber\\
\eea
where the extended counterterms
$\deltaZ{1}{\gamma_i}{}{}
+
\deltaZtilde{1}{\gamma_i}{}{}$ 
guarantee the consistent subtraction of UV poles
in $\numdim=4$ dimensions \refeq{eq:4dimsubdiagB2}.
The $\ntilde$-contributions stemming from 
subdivergences, see \refeq{eq:ddimsubdiagD2},  are reconstructed by 
\bea
\label{eq:twoloopratstC}
\bfKtildesub\, \amp{2}{\Gamma}{}{}&=&
\sum_{i=1}^3
(\bfKtilde\, \amp{1}{\gamma_i}{}{})\cdot \amp{1}{\Gamma/\gamma_i}{}{}
\,=\,
\sum_{i=1}^3
\int\rd\barq_i\,
\ratamp{1}{\gamma_i}{\alpha_i}{q_i}\,
\frac{\calN^{(i)}_{\alpha_i}(q_i)}{\calD{i}}
\,.
\eea
For what concerns the subtraction of local divergences, 
up to negligible terms of $\ord(\eps)$ the 
$\bfKloc$ operator in \refeq{eq:twoloopratstA} is equivalent to
its $\dendim$-dimensional version, \ie 
\bea
\label{eq:twoloopratstD}
\bfKloc\, \ampbar{2}{\Gamma}{}{}
&=&
\bfKloc\, \amp{2}{\Gamma}{}{} 
\,=\,
{}-\deltaZ{2}{\Gamma}{}{}\,,
\eea
where $\deltaZ{2}{\Gamma}{}{}$ is the usual $\msbar$ two-loop counterterm.
The remaining $\bfKtildeloc$ operator
describes the $\ntilde$-contributions stemming from 
local two-loop divergences and
is implicitly defined through \refeq{eq:twoloopratstA} as
\bea
\label{eq:twoloopratstE}
\bfKtildeloc\,\amp{2}{\Gamma}{}{}
&=&
\left(1-\bfKsub\right)\ampbar{2}{\Gamma}{}{}
-\left(1-\bfKsub+\bfKtildesub\right) \amp{2}{\Gamma}{}{}\,.
\eea
As demonstrated in the next section, such $\ntilde$-contributions can be 
reconstructed through process-independent 
counterterms,
\bea
\label{eq:twoloopratstF}
\bfKtildeloc\,\amp{2}{\Gamma}{}{}
&=&
\ratamp{2}{\Gamma}{}{}\,,
\eea
which can be computed once and for all in terms of tadpole integrals.
This implies that the $\ratamp{2}{\Gamma}{}{}$ counterterms are 
polynomials in the external momenta and can be described at the level of the
Lagrangian in terms
of local operators.

\begin{figure}[t]
\begin{center}
\bea
\hspace{-5mm}
&&\bfR\,
\left[\;
\vcenter{\hbox{\includegraphics[width=\diawidth]{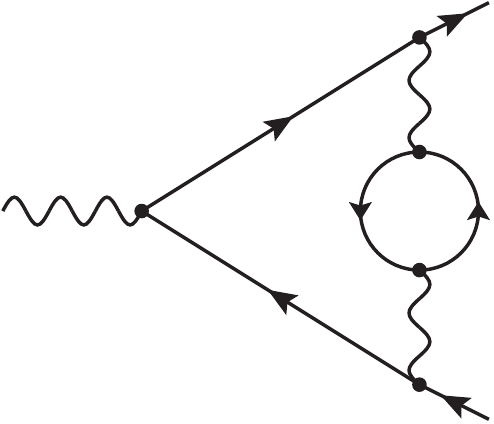}}}\;\;\right]_{\numdim\,=\,\dendim}
\hspace{-3mm}
\,=\, \nonumber\\[5mm]
\hspace{-5mm}&&=\,
\left[\;
\vcenter{\hbox{\includegraphics[width=\diawidth]{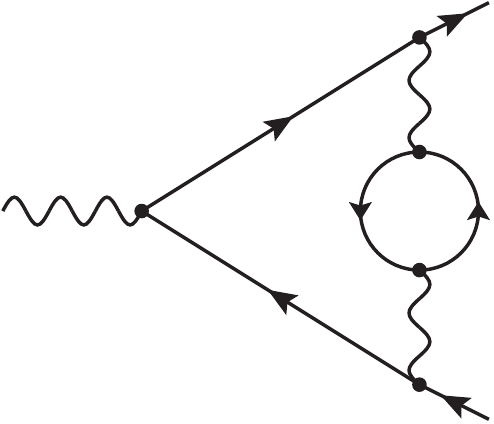}}}  
\;\;+\;
\vcenter{\hbox{\includegraphics[width=\diawidth]{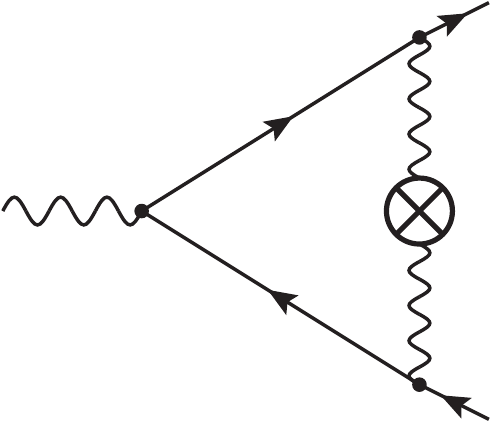}}}{
\left(\deltaZ{1}{\gamma_i}{}{}+ \deltaZtilde{1}{\gamma_i}{}{} 
+\ratamp{1}{\gamma_i}{}{}\right)
} 
\;+\;
\vcenter{\hbox{\includegraphics[width=\diawidth]{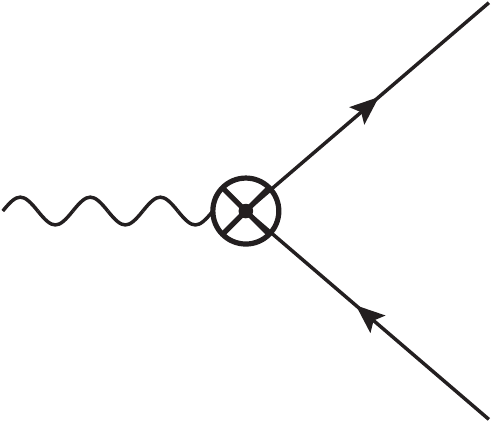}}} \hspace{-5mm}
\left(\deltaZ{2}{\Gamma}{}{} 
+\ratamp{2}{\Gamma}{}{}\right)
\;\right]_{\numdim\,=\,4}
\nonumber
\eea%s
\end{center}
\caption{Graphical representation of the master formula \refeq{eq:twoloopratstA}, see also
\refeq{eq:masterformula}, for a globally divergent two-loop QED diagram with a single subdivergence.} 
\label{fig:twoloopratQED}
\end{figure}

The master formula \refeq{eq:twoloopratstA} is equivalent to 
\refeq{eq:masterformula} and can be written more 
explicitly in terms of loop integrals as
\bea
\label{eq:twoloopratstG}
&&\bfR\,\ampbar{2}{\Gamma}{}{} 
\,=\,
\int\rd\barq_1
\int \rd\barq_2\,
\frac{\calN(q_1,q_2)}
{\calD{1}\,\calD{2}\,\mathcal{D}^{(3)}(-\barq_1-\barq_2)}
\nonumber\\&&\qquad{}
+\sum_{i=1}^3
\int\rd\barq_i\,
\left[
\deltaZ{1}{\gamma_i}{\alpha_i}{q_i}
+\deltaZtilde{1}{\gamma_i}{\alpha_i}{\tilde q_i}
+\ratamp{1}{\gamma_i}{\alpha_i}{q_i}
\right]
\frac{\calN^{(i)}_{\alpha_i}(q_i)}{\calD{i}}
\,+\,\left(\deltaZ{2}{\Gamma}{}{}
+\ratamp{2}{\Gamma}{}{}\right)\,.
\nonumber\\
\eea 
Note that the numerator of the two-loop integral on the rhs is strictly 
four-dimensional, while the presence of $1/\eps$ and $\tildeqidx{}{2}/\eps$
poles in $\deltaZ{1}{\gamma_i}{}{}$ and $\deltaZtilde{1}{\gamma_i}{}{}$
requires the evaluation of one-loop 
integrals of type 
\bea
T_{N,s}^{\mu_1\cdots \mu_r}
&=&
\int\rd\barq_1\, \frac{
(\tildeqidx{}{2})^s\,
 q_1^{\mu_1}\cdots  q_1^{\mu_r}}
{D_0(\barq_1)\cdots D_{N-1}(\barq_1)}
\eea
up to $\ord(\eps^{1})$, where the 
$q_1^{\mu_i}$ loop momenta in the numerator are four-dimensional, while
the additional factor $\tildeqidx{}{2}$ has power $s=0$ or $1$.

Explicit results for all relevant UV and rational 
counterterms in QED are presented in \refse{sec:qedres}, and the
structure of the above master formula
for a two-loop QED diagram
is illustrated in \reffi{fig:twoloopratQED}.

\subsection{Proof and recipe for the calculation of rational terms}
\label{se:proof}

As pointed out in the previous section, the master formula \refeq{eq:twoloopratstA} 
should be regarded as implicit definition of the
$\bfKtildeloc$ operator, which is explicitly defined in \refeq{eq:twoloopratstE}.
By construction $\bfKtildeloc$ embodies the two-loop
$\ntilde$-contributions that remain after subtraction of 
all UV divergences and of the rational parts stemming from 
one-loop subdivergences.
In the following we demonstrate that such 
$\ntilde$-contributions can be reduced to process-independent local counterterms
$\ratamp{2}{\Gamma}{}{}$
as anticipated in~\refeq{eq:twoloopratstF}. 
The proof consists of two parts, which 
deal with diagrams with 
$X(\Gamma)<0$ and $X(\Gamma)\ge 0$, respectively.
We also provide an explicit recipe to calculate
the two-loop rational counterterms $\ratamp{2}{\Gamma}{}{}$
by means of tadpole integrals.

\subsubsection{Diagrams with $X(\Gamma)<0$}
\label{se:2loopnonglobdivproof}
We first consider generic two-loop diagrams $\Gamma$ 
with $X(\Gamma)<0$.
This implies that $\Gamma$ is free from local divergences, \ie
\bea
\label{eq:noglobdivA0}
\bfKloc\, \ampbar{2}{\Gamma}{}{}
&=&
0\,.
\eea
Thus only subdivergences need to be renormalised, \ie
\bea
\label{eq:noglobdivA00}
\bfR \, \ampbar{2}{\Gamma}{}{} &=&
\ampbar{2}{\Gamma}{}{} -\bfKsub\,\ampbar{2}{\Gamma}{}{}\,.
\eea
Since for $X(\Gamma)<0$ at most one subdiagram can be UV divergent (see
\refse{se:twolooppowcount}),
using \refeq{eq:twolooprenB} and \refeq{eq:subdiagnotB},
we can write\footnote{
The following identity can be written more explicitly 
as
\bea
\label{eq:subdiagnotA2}
\bfR\, \ampbar{2}{\Gamma}{}{} 
&=&
\int \rd\barq_i\,
\int\rd\barq_j\,
\frac{\bar\calN(\barq_i,\barq_j)}
{\calD{i}\,\calD{j}\,\calD{k}}
+
\int \rd\barq_i\,
\deltaZ{1}{\gamma_i}{\bar\alpha_i}{\barq_i}\,
\frac{\barN^{(i)}_{\bar\alpha_i}(\barq_i)}{\calD{i}}
\nonumber\\
&=&
\int \rd\barq_i\,
\left[\int\rd\barq_j \frac{\bar\Gamma^{\bar\alpha_i\bar\alpha_j\bar\alpha_k}(\barq_i,\barq_j,\barq_k)
\barN^{(j)}_{\bar\alpha_j}(\barq_j)\barN^{(k)}_{\bar\alpha_k}(\barq_k)}{\calD{j}\calD{k}}
+\deltaZ{1}{\gamma_i}{\bar\alpha_i}{\barq_i} \right] 
\frac{\barN^{(i)}_{\bar\alpha_i}(\barq_i)}{\calD{i}}\,,
\eea
where the integral representations in 
\refeq{eq:subdiagnotA}--\refeq{eq:subdiagnotB}
and \refeq{eq:twolooprenB} are used
with $\barq_k=-\barq_i-\barq_j$,  and
$i|jk$ is a partition of $123$.
In the above integral representation,
the identities \refeq{eq:noglobdivZA} and \refeq{eq:noglobdivZB} correspond,
respectively, to 
\bea
\label{eq:subdiagnotB2}
\frac{\barN^{(i)}_{\bar\alpha_i}(\barq_i)}{\calD{i}}
&=&
\frac{\calN^{(i)}_{\alpha_i}(q_i)}{\calD{i}}+\ord(\eps)\,,
\eea
and
\bea
\label{eq:subdiagnotC2}
\bigg[\dots\bigg]&=&
\int\rd\barq_j \frac{\Gamma^{\alpha_i\alpha_j\alpha_k}(q_i,q_j,q_k)
\calN^{(j)}_{\alpha_j}(q_j)\calN^{(k)}_{\alpha_k}(q_k)}{\calD{j}\calD{k}} 
+\deltaZ{1}{\gamma_i}{\alpha_i}{q_i} 
+\deltaZtilde{1}{\gamma_i}{\alpha_i}{\tilde q_i}
+\ratamp{1}{\gamma_i}{\alpha_i}{q_i}
+\ord(\eps)\,,\qquad
\eea
where $[\,\dots]$ refers to the term between square brackets in \refeq{eq:subdiagnotA2}.
Note that on the rhs of \refeq{eq:subdiagnotB2} and \refeq{eq:subdiagnotC2}
the loop numerators of $\Gamma/\gamma_i$ and $\gamma_i$, including 
the connecting multi-index $\alpha_i$, 
are projected to four dimensions.
}
\bea
\label{eq:noglobdivA}
\bfR \, \ampbar{2}{\Gamma}{}{} &=&
\ampbar{2}{\Gamma}{}{} -
\left(\bfK\ampbar{1}{\gamma_i}{}{}\right)\cdot \ampbar{1}{\Gamma/\gamma_i}{}{}
\,=\,
\left(\ampbar{1}{\gamma_i}{}{}-\bfK\ampbar{1}{\gamma_i}{}{}\right)
\cdot \ampbar{1}{\Gamma/\gamma_i}{}{}\,,
\eea
where we assume that the subdiagram $\gamma_i$ can 
be divergent or non-divergent, in which case
$\bfK\ampbar{1}{\gamma_i}{}{}=0$, while the two remaining subdiagrams are
free from divergences.
The two factors on the rhs of \refeq{eq:noglobdivA}
can be related to corresponding four-dimensional quantities 
using 
\bea
\label{eq:noglobdivZA}
\ampbar{1}{\Gamma/\gamma_i}{}{}
&=&
\amp{1}{\Gamma/\gamma_i}{}{} +\ord(\eps)\,,
\eea
and the identity \refeq{eq:subdivP} for the $\gamma_i$ subdiagram,
which corresponds to 
\bea
\label{eq:noglobdivZB}
\ampbar{1}{\gamma_i}{}{}-\bfK\ampbar{1}{\gamma_i}{}{}
&=&
\amp{1}{\gamma_i}{}{}-\bfK\amp{1}{\gamma_i}{}{}+\bfKtilde\amp{1}{\gamma_i}{}{}\,
+\ord(\eps)\,,
\eea
where the dependence on the loop momentum $q_i$ and the connecting multi-index 
$\alpha_i$ is kept implicit,
and contributions of order $\tildeqidx{i}{2}/\eps$ are consistently taken into
account through the $\bfK$ operator as detailed in 
\refeq{eq:4dimsubdiagB2}--\refeq{eq:ztildeformA}.
The identities \refeq{eq:noglobdivZA}--\refeq{eq:noglobdivZB} 
can be directly applied on the rhs of \refeq{eq:noglobdivA}
neglecting all terms of $\ord(\eps)$
since the renormalised subdiagram $\gamma_i$ and its complement 
$\Gamma/\gamma_i$ are both free from UV singularities.
This results into
\bea
\label{eq:noglobdivB2}
\bfR\,\ampbar{2}{\Gamma}{}{}
&=&
\left(\amp{1}{\gamma_i}{}{}-\bfK\amp{1}{\gamma_i}{}{}+\bfKtilde\amp{1}{\gamma_i}{}{}\right)
\cdot \amp{1}{\Gamma/\gamma_i}{}{}\,,
\eea
which can be rewritten in terms of the subtraction 
operators defined in \refeq{eq:twoloopratstB}--\refeq{eq:twoloopratstC}
as
\bea
\label{eq:noglobdivDA}
\bfR\,\ampbar{2}{\Gamma}{}{}
&=&
\amp{2}{\Gamma}{}{}+
\left(\bfKtildesub-\bfKsub\right) \amp{2}{\Gamma}{}{}\,.
\eea
This equation demonstrates the validity of the master formula \refeq{eq:twoloopratstA}
for the case $X(\Gamma)<0$ and shows that in this case neither 
local divergences nor $\ntilde$-contributions of type
\refeq{eq:twoloopratstE}--\refeq{eq:twoloopratstF} occur, \ie
\bea
\label{eq:noglobdivD}
X(\Gamma)<0
\qquad\Rightarrow \qquad
\bfKloc\,\amp{2}{\Gamma}{}{}\,=\,\deltaZ{2}{\Gamma}{}{} \,=\,0
\quad\mbox{and}\quad
\bfKtildeloc\,\amp{2}{\Gamma}{}{} \,=\,\ratamp{2}{\Gamma}{}{} \,=\,0
\,.
\eea
Thus the genuine two-loop $\ntilde$-terms
\refeq{eq:twoloopratstE}--\refeq{eq:twoloopratstF}
occur only in 
the finite set of diagrams that involve a local divergence.
Note that \refeq{eq:noglobdivDA} and \refeq{eq:noglobdivD}
hold irrespectively of the presence of subdivergences, \ie also for 
finite two-loop diagrams.
Moreover, due to the linearity of the various $\bfK$ operators,
the above identities are applicable to 
full diagrams, combinations of diagrams, or 
single pieces of individual diagrams.

\begin{figure}[t]
\begin{center}
\bea
\bfR\;\left[\;\,
\vcenter{\hbox{\includegraphics[height=\diaheight]{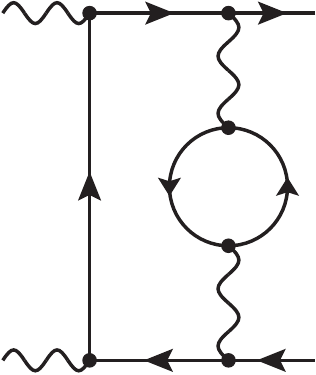}}}
\;\;\right]_{\numdim\,=\,\dendim}
\hspace{-2mm}
&=& 
\left[ \;\,
\vcenter{\hbox{\includegraphics[height=\diaheight]{QEDfptIILoop}}}
\;\;+\;\;
\vcenter{\hbox{\includegraphics[height=\diaheight]{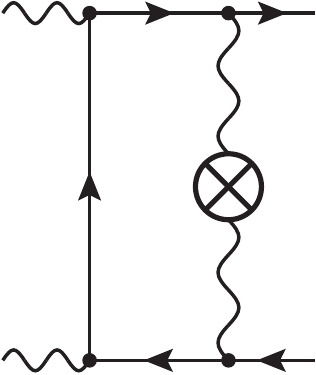}}} 
\Big(\deltaZ{1}{\gamma}{}{}
+
\deltaZtilde{1}{\gamma}{}{} 
+
\ratamp{1}{\gamma}{}{}
\Big)
\right]_{\numdim\,=\,4}  
\nonumber
\eea
\end{center}
\caption{
Graphical representation of the master formula \refeq{eq:twoloopratstA}, see also
\refeq{eq:masterformula}, for a two-loop QED diagram where $X(\Gamma)<0$ and
thus $\deltaZ{2}{\gamma}{}{} = \ratamp{2}{\gamma}{}{} =0$.
} 
\label{fig:twoloopnoglobdiv}
\end{figure}

A schematic representation of equation~\refeq{eq:noglobdivDA}
for the case of 
a two-loop QED diagram with $X(\Gamma)<0$ 
and a subdivergence is shown in \reffi{fig:twoloopnoglobdiv}.

\subsubsection{Diagrams with $X(\Gamma)\ge 0$}
\label{se:2loopglobdivproof}

In the following we consider two-loop diagrams $\Gamma$ with $X(\Gamma)\ge 0$,
and we prove that $\bfKtildeloc\amp{2}{\Gamma}{}{}$
can be cast in the form of tadpole integrals 
and corresponds to a local counterterm.
As detailed below,  the proof is based on the
splitting of $\Gamma$ into two parts,
\bea
\label{eq:tadexptwoloopC}
\ampbar{2}{\Gamma}{}{} &=& 
\ampbar{2}{\Gamma_\tad}{}{} 
+
\ampbar{2}{\Gamma_\rem}{}{}\,,
\eea
where $\Gamma_\tad$ embodies the entire globally divergent 
part of $\Gamma$ in the form of pure tadpole integrals, while
$\Gamma_\rem$ is not globally divergent, \ie
\bea
\label{eq:globdivsplit}
X(\Gamma_\tad)&=&X(\Gamma)\,\ge\, 0
\qquad\mbox{and}\qquad
X(\Gamma_\rem)\,<\,0\,.
\eea
This allows us to apply 
\refeq{eq:noglobdivD} to $\Gamma_\rem$ and to conclude that 
\bea
\label{eq:tadexptwoloopG}
\bfKtildeloc\, \amp{2}{\Gamma_\rem}{}{}
\,=\,0
\qquad\mbox{and}\qquad
\bfKtildeloc\, \amp{2}{\Gamma}{}{}
\,=\,
\bfKtildeloc\, \amp{2}{\Gamma_\tad}{}{}
\,,
\eea
where the second relation follows form the 
linearity of the $\bfKtildeloc$ operator.

The splitting \refeq{eq:tadexptwoloopC}
is implemented by means of the tadpole decomposition 
introduced in \refse{se:tadpoledec}.
Specifically, along each chain $\calC_i$ of the two-loop diagram
we apply an exact decomposition,
\bea
\label{eq:tadexptwoloopG2}
\bfS^{(i)}_{X_i}+\bfF^{(i)}_{X_i}&=&1\,,
\eea
where the operators $\bfS^{(i)}_{X_i}$ and $\bfF^{(i)}_{X_i}$
act exclusively on the denominators depending on the loop momentum
$\barq_i$ and are defined similarly as in 
\refeq{eq:dentadexpC}--\refeq{eq:chaintadexpB}. In particular,
the $\bfS^{(i)}_{X_i}$ operator generates the tadpole expansion
\bea
\label{eq:tadexptwoA}
\bfS^{(i)}_{X_i}\lb 
\frac{1}{\calD{i}}\rb
&=& 
\sum_{\sigma=0}^{X_i}
\frac
{\Delta_i^{(\sigma)}(\barq_i)}
{\left(\barqidx{i}{2}-M^2\right)^{N_i+\sigma}}\,,
\eea
where
\bea
\label{eq:tadexptwoB}
{\Delta_i^{(\sigma)}(\barq_i)}
&=& 
\sum_{\sigma_0=0}^\sigma
\ldots
\sum_{\sigma_{N_i-1}=0}^\sigma
\prod_{a=0}^{N_i-1}
\Big[\Delta_{ia}(\barq_i)\Big]^{\sigma_a}
\Bigg|_{\sigma_0+\dots+ \sigma_{N_i-1} = \sigma}
\eea
with
\bea
\label{eq:tadexptwoC}
\Delta_{ia}(\barq_i)&=& 
-p_{ia}^2-2 \barq_i\cdot p_{ia}+m_{ia}^2-M^2\,.
\eea
In practice $\bfS^{(i)}_{X_i}$ turns all propagators along the chain 
$\calC_i$ into tadpoles including subleading 
UV contributions up to a certain 
relative
order $1/q_i^{X_i}$, 
while $\bfF^{(i)}_{X_i}$ collects all remnant terms, which are suppressed by
order $1/q_i^{X_i+1}$ or higher.
Therefore, each $\bfF^{(i)}_{X_i}$ operator reduces 
the degree of divergence of all (sub)diagrams that involve
the chain $\calC_i$ by $X_i+1$.
More explicitly, for the global degree of divergence
\bea
\label{eq:divredA}
X\lb\bfF^{(i)}_{X_i}\,\Gamma\rb
&\le& X(\Gamma)-\lb X_i+1\rb 
\,,
\eea
and for the degree of divergence 
of the subdiagrams involving the chains 
$\calC_i\calC_j$ and
$\calC_i\calC_k$,
\bea
\label{eq:divredB}
X_{il}\lb\bfF^{(i)}_{X_i}\,\Gamma\rb
&\le& X_{il}(\Gamma)-\lb X_i+1\rb 
\qquad\mbox{for}\quad l=j,k\,,
\eea
where $i|jk$ is a partition of $123$, while
$X_{ik}(\Gamma)=X(\gamma_j)$ and $X_{ij}(\Gamma)=X(\gamma_k)$
are defined in \refeq{eq:Xijdef}.
Based on \refeq{eq:divredA}--\refeq{eq:divredB}
the order of the tadpole decompositions 
along the various chains
is chosen as
\bea
X_i &=&
X_i(\Gamma) \,=\,
\text{max}\left\{ X(\Gamma), X_{ij}(\Gamma),
X_{ik}(\Gamma)\right\}\,,
\label{eq:powercountingDirectImp}
\eea
which guarantees that 
\bea
\label{eq:divredC}
X\lb\bfF^{(i)}_{X_i}\, \Gamma\rb
&<&0\,,
\eea
and
\bea
\label{eq:divredD}
X_{ij}\lb\bfF^{(i)}_{X_i}\Gamma\rb<0\,,\quad
X_{ik}\lb\bfF^{(i)}_{X_i}\, \Gamma\rb
\,<\,0\,.
\eea
Thus, the remnant part $\bfF^{(i)}_{X_i}\,\Gamma$ that results from the 
decomposition \refeq{eq:tadexptwoloopG2}
of a {\it single} chain  $\calC_i$
is completely free from global two-loop 
divergences and contains only 
UV divergences stemming from the subdiagram
that does not involve  the chain $\calC_i$,
\ie the $\gamma_i$  subdiagram.
Vice versa, the tadpole part 
$\bfS^{(i)}_{X_i}\,\Gamma$ 
contains the entire 
globally divergent part of $\Gamma$
as well as the 
UV divergences of the subdiagrams $\gamma_j$ and $\gamma_k$, which 
involve the chains $\calC_i\calC_j$ and $\calC_i\calC_k$, respectively.

The decomposition of all three chains
can be expressed as
\bea
\label{eq:tadexptwoloopB}
\ampbar{2}{\Gamma}{}{} 
&=& 
\left(\bfS^{(1)}_{X_1}+\bfF^{(1)}_{X_1}\right)
\left(\bfS^{(2)}_{X_2}+\bfF^{(2)}_{X_2}\right)
\left(\bfS^{(3)}_{X_3}+\bfF^{(3)}_{X_3}\right)
\ampbar{2}{\Gamma}{}{}\,,
\eea
or, more explicitly,
\bea
\label{eq:tadexptwoloopB1}
\ampbar{2}{\Gamma}{}{} 
&=& 
\int\rd\barq_1
\int \rd\barq_2\,
\bar\calN(\barq_1,\barq_2)
\prod_{i=1}^3
\left[
\left(\bfSX{i}+\bfFX{i}\right)\,\frac{1}{\calD{i}}
\right]_{q_3=-q_1-q_2}
\,.
\eea
Expanding the rhs of \refeq{eq:tadexptwoloopB}
results into eight different combinations of $\bfS$ and $\bfF$
operators, which can be grouped into two 
contributions that correspond to the tadpole and
remnant parts introduced in \refeq{eq:tadexptwoloopC},
\bea
\label{eq:tadexptwoloopD}
\ampbar{2}{\Gamma_\tad}{}{} 
&=&
\bfS^{(1)}_{X_1}
\bfS^{(2)}_{X_2}
\bfS^{(3)}_{X_3}
\ampbar{2}{\Gamma}{}{}\,,
\\
\label{eq:tadexptwoloopE}
\ampbar{2}{\Gamma_\rem}{}{} 
&=&
\bfF^{(1)}_{X_1}
\bfF^{(2)}_{X_2}
\bfF^{(3)}_{X_3}
\ampbar{2}{\Gamma}{}{} 
+
\sum_{i=1}^3
\bfS^{(i)}_{X_i}
\bfF^{(j)}_{X_j}
\bfF^{(k)}_{X_k}
\ampbar{2}{\Gamma}{}{} 
+
\sum_{i=1}^3 
\bfF^{(i)}_{X_i}
\bfS^{(j)}_{X_j}
\bfS^{(k)}_{X_k}
\ampbar{2}{\Gamma}{}{} 
\,.
\eea
In \refeq{eq:tadexptwoloopD}
all three chains are replaced by their tadpole expansion, \ie
$\Gamma_\tad$ consists only of tadpole
integrals. In contrast, the remaining seven combinations in 
\refeq{eq:tadexptwoloopE} involve at least one 
$\bfF$ operator. Thus  \refeq{eq:divredC} guarantees that
the $\Gamma_\rem$ part is  free form global divergences and does not 
contribute to $\bfKtildeloc\, \amp{2}{\Gamma}{}{}$,
as anticipated in \refeq{eq:globdivsplit}--\refeq{eq:tadexptwoloopG}.

Combining \refeq{eq:tadexptwoloopG} with
\refeq{eq:twoloopratstE} we obtain
\bea
\label{eq:tadexptwoloopH}
\bfKtildeloc\, \amp{2}{\Gamma}{}{}
&=&
\left(1-\bfKsub\right)\ampbar{2}{\Gamma_\tad}{}{}
-\left(1-\bfKsub+\bfKtildesub\right) \amp{2}{\Gamma_\tad}{}{}\,.
\eea
Here the interplay between the tadpole expansion \refeq{eq:tadexptwoloopD} 
and the $\bfKsub$ operators results into 
tadpole terms like
\bea
\label{eq:tadexptwoloopI}
\bfKsub\, \ampbar{2}{\Gamma_\tad}{}{} &=&
\sum_{i=1}^3
(\bfK\, \ampbar{1}{\gamma_{\tad,i}}{}{})\cdot
\ampbar{2}{\Gamma_\tad/\gamma_{\tad,i}}{}{}
\,=\,
\sum_{i=1}^3
\left(\bfK\, \bfS^{(j)}_{X_j} \bfS^{(k)}_{X_k}
\ampbar{1}{\gamma_i}{}{}\right)\cdot
\left(\bfS^{(i)}_{X_i} \ampbar{2}{\Gamma/\gamma_i}{}{}\right)
\nonumber\\
&=&
\sum_{i=1}^3
\left(\bfK\, \ampbar{1}{\gamma_i}{}{}\right)\cdot
\left(\bfS^{(i)}_{X_i} \ampbar{2}{\Gamma/\gamma_i}{}{}\right)\,,
\eea
where the second identity is guaranteed by the fact that 
the $\bfS$ operators in \refeq{eq:tadexptwoloopD}
factorise, in the sense that each 
$\bfS^{(i)}_{X_i}$ acts only on the corresponding chain
$\calC_i$.
The third identity is based on \refeq{eq:divredD},
which guarantees that the tadpole expansions
$\bfS^{(j)}_{X_j}$ and $ \bfS^{(k)}_{X_k}$
capture the full UV divergent part of the subdiagram $\gamma_i$
containing the chains $\calC_j$ and $\calC_k$.
Using similar identities for the terms 
$\bfKsub\, \amp{2}{\Gamma_\tad}{}{}$  and
$\bfKtildesub\, \amp{2}{\Gamma_\tad}{}{}$ in  
\refeq{eq:tadexptwoloopH} we arrive at 
\bea
\label{eq:tadexptwoloopK}
\bfKtildeloc\,\amp{2}{\Gamma}{}{}
&=&
\left[
\bfSall
\,\ampbar{2}{\Gamma}{}{}
-\sum_{i=1}^3
\left(\bfK \ampbar{1}{\gamma_i}{}{}\right)\cdot
\left(\bfS^{(i)}_{X_i} \ampbar{2}{\Gamma/\gamma_i}{}{}\right)
\right]
\nonumber\\
&&{}-
\left[
\bfSall
\,\amp{2}{\Gamma}{}{}
-\sum_{i=1}^3
\left(\bfK\,\amp{1}{\gamma_i}{}{}-\bfKtilde\, \amp{1}{\gamma_i}{}{}\right)\cdot
\left(\bfS^{(i)}_{X_i} \amp{2}{\Gamma/\gamma_i}{}{}\right)
\right]\,,
\eea
which can be written more explicitly as
\bea
\label{eq:tadexptwoloopL}
\ratamp{2}{\Gamma}{}{}
&=&
\int\rd\barq_1
\int \rd\barq_2
\left[\bar\calN(\barq_1,\barq_2)- \calN(q_1,q_2)\right]
\left[\prod_{i=1}^3 \bfSX{i}\,\frac{1}{\calD{i}}
\right]_{q_3=-q_1-q_2}
\nonumber\\
&&\hspace{-4mm}+\sum_{i=1}^3
\int\rd\barq_i\,
\left[
\deltaZ{1}{\gamma_i}{\bar\alpha_i}{\barq_i}\,
\bar\calN^{(i)}_{\bar\alpha_i}(\barq_i)
-
\left(\deltaZ{1}{\gamma_i}{\alpha_i}{q_i}
+\deltaZtilde{1}{\gamma_i}{\alpha_i}{\tilde q_i}+\ratamp{1}{\gamma_i}{\alpha_i}{q_i}\right)
\calN^{(i)}_{\alpha_i}(q_i)
\right]
\nonumber\\
&&\hspace{16mm}{}\times\bfS^{(i)}_{X_i}
\left(\frac{1}{\calD{i}}\right)
\,.
\eea 
The identities \refeq{eq:tadexptwoloopK}--\refeq{eq:tadexptwoloopL}
represent the master formulas for the 
calculation of
the $\ratamp{2}{\Gamma}{}{}$ counterterms in terms of tadpole integrals.
Moreover, the structure of these formulas provides 
insights into the general properties of the $\ratamp{2}{\Gamma}{}{}$
counterterms.
In particular,
from the form of \refeq{eq:tadexptwoloopL} 
and \refeq{eq:tadexptwoA}--\refeq{eq:tadexptwoC} 
it is evident that such counterterms
are polynomials in the
external momenta  $\{p_{ia}\}$ 
and internal masses $\{m_{ia}\}$.
With other words, the 
$\ratamp{2}{\Gamma}{}{}$ counterterms
correspond to local operators at the Lagrangian level,
and at the level of scattering amplitudes they 
result into rational functions of the 
kinematic invariants.

The various tadpole expansions in \refeq{eq:tadexptwoloopL}
give rise to terms depending on the 
auxiliary mass scale $M^2$. However, 
this dependence cancels in $\ratamp{2}{\Gamma}{}{}$.
This is guaranteed by the fact that the
tadpole decomposition~\refeq{eq:tadexptwoloopB} is exact,
and thus independent of $M^2$, while
the contribution of the amputated remnant \refeq{eq:tadexptwoloopE}
to $\delta\calR_2$ vanishes.
This implies that $\delta \calR_2$ counterterms are also independent of the 
renormalisation scale $\mu$,  
since such dependence could arise only in the form of 
logarithms of $M^2/\mu^2$ in the tadpole integrals\footnote{In general the
rational terms depend on the ratio between the regularisation and renormalisation scales.  
However these two scales are set equal to each other in this paper.}
on the rhs of \refeq{eq:tadexptwoloopL}.

\begin{figure}[t]
\begin{center}
\bea
\bfKtildeloc\,\left[\,\,
\vcenter{\hbox{\includegraphics[width=\diawidth]{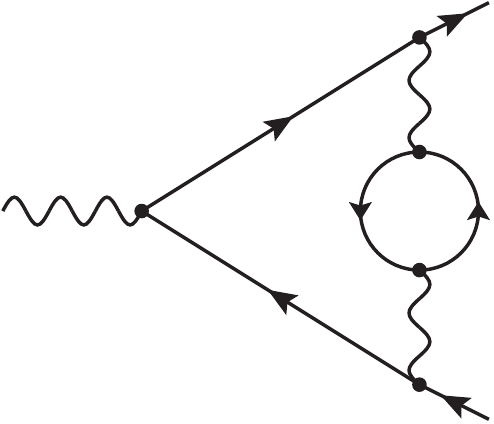}}}  
\;\;\right]_{\numdim\,=\,\dendim}
\;=\;\;\;
\left[ \;\,
\bfSall\,
\vcenter{\hbox{\includegraphics[width=\diawidth]{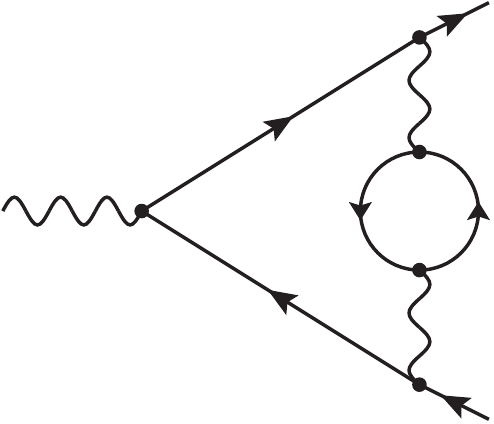}}}  
\;\;+\;\; \bfS^{(1)}_{X_1}\,
\vcenter{\hbox{\includegraphics[width=\diawidth]{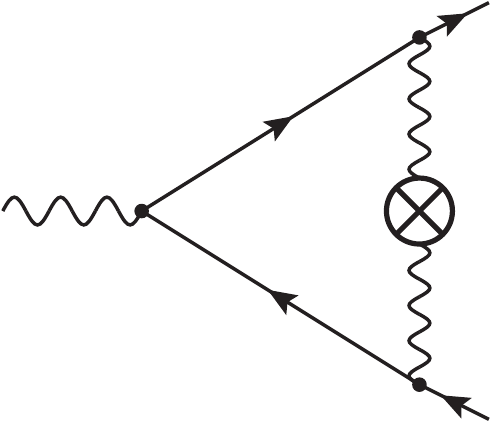}}}
\deltaZ{1}{\gamma_1}{}{}
\right]_{\numdim\,=\,\dendim} && 
\nonumber\\[6mm]
{}-\;\;
\left[\,
\bfSall\,
\vcenter{\hbox{\includegraphics[width=\diawidth]{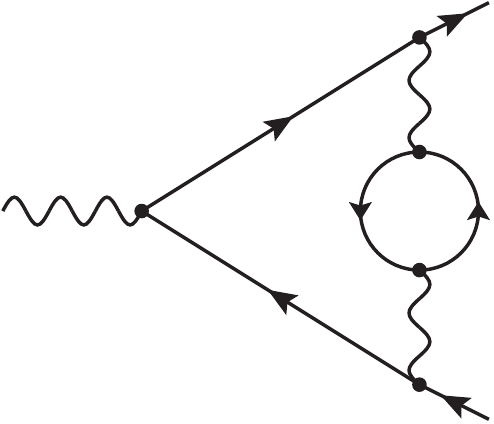}}}  \;\; + \;\; \bfS^{(1)}_{X_1}\,
\vcenter{\hbox{\includegraphics[width=\diawidth]{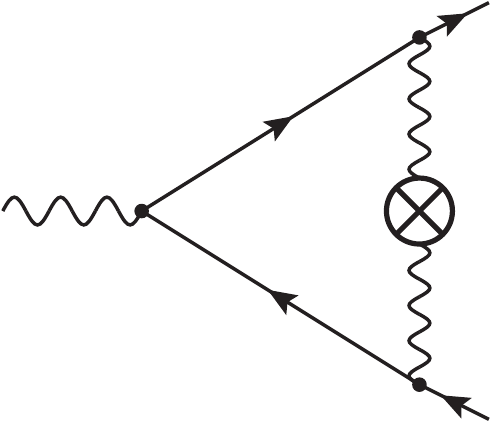}}}
\Big({ \deltaZ{1}{\gamma_1}{}{}} 
+ { \deltaZtilde{1}{\gamma_1}{}{} } 
+ { \ratamp{1}{\gamma_1}{}{}}\Big) 
\right]_{D=4} &&
\nonumber
\eea
\end{center}
\caption{Graphical representation of the master 
formula~\refeq{eq:tadexptwoloopK}
for the derivation of two-loop rational counterterms 
for the case of a globally divergent two-loop diagram with a single
divergent subdiagram $\gamma_1$. The $\bfS^{(i)}_{X_i}$ operators perform tadpole
expansions along the corresponding chains $\calC_i$, and
the subtracted one-loop contributions involve 
a single tadpole expansion
along the chain $\calC_1$ associated with
the complement $\Gamma/\gamma_1$ of the divergent subdiagram.
} 
\label{fig:twolooprat}
\end{figure}

The master formula \eqref{eq:tadexptwoloopL}
can be optimised in various ways.
For instance, the number of tadpole integrals to 
be computed can be significantly reduced by applying a 
strict power counting in $1/q_i$ such that 
all terms of relative order higher than $1/q_i^{X_i}$ are shifted
from the $\bfS_{X_i}^{(i)}$ operators
to the $\bfF_{X_i}^{(i)}$ remnants.
Moreover, the fact that the 
resulting $\ratamp{2}{\Gamma}{}{}$ terms are 
homogenous polynomials of degree
$X(\Gamma)$ in 
$\{p_{ia}, m_{ia}\}$
allows one to discard all 
terms of different order  at the integrand level.
The results presented in \refse{sec:qedres} have been obtained 
by selecting the terms of order
$X(\Gamma)$ in $\{p_{ia}, m_{ia}\}$
and discarding also all irrelevant $M^2$-dependent terms in the loop numerators. 
This can be achieved by omitting all $M^2$-contributions in
\refeq{eq:tadexptwoB}
and then reconstructing the correct $M^2$-dependence 
through auxiliary one-loop counterterms along the lines of~\cite{Misiak:1994zw,beta_den_comp,
Zoller:2014xoa}.
Results obtained in this way have been validated against a naive
implementation of the tadpole expansions as described in
\refeq{eq:tadexptwoA}--\refeq{eq:tadexptwoB}.
More details on the implementation of the tadpole expansion and 
its optimisations will be discussed in a forthcoming paper.

The master formula \refeq{eq:tadexptwoloopK}--\refeq{eq:tadexptwoloopL}
for the calculation of two-loop rational counterterms
is illustrated 
in \reffi{fig:twolooprat} for the case of a two-loop QED diagram with a single
subdivergence. 

\section{Two-loop rational terms in QED} \label{sec:qedres}

\newcommand{\shiftleft}{\!\!\!\!\!\!\!\!\!\!\!\!\!\!\!\!\!\!\!}

As a first application of the method introduced in \refse{se:irredtwoloop}
we have derived the full set of two-loop rational terms 
$\delta\calR_2$
in QED
in the $\msbar$ scheme.
To this end we have applied the master formula \refeq{eq:tadexptwoloopL}
to the full set of globally divergent Feynman diagrams 
for the various 2-, 3-, and 4-point 1PI functions in QED.
For convenience, let us summarise the various ingredients that are required for
the implementation of \refeq{eq:tadexptwoloopL}.
As defined in \refeq{eq:twoloopnotA1}--\refeq{eq:rttwoloopFF},
the function $\bar\calN(\barq_1,\barq_2)$ 
is the numerator of the two-loop diagram at hand in $\dendim$ dimensions,
$\calN(q_1,q_2)$ represents its four-dimensional 
projection, and $\calD{i}$ are the associated chains of denominators.
The functions $\bar\calN^{(i)}_{\bar\alpha_i}(\barq_i)$
and $\calN^{(i)}_{\alpha_i}(q_i)$ represent,
respectively, the $\dendim$- and four-dimensional 
numerator of the complement $\Gamma/\gamma_i$ 
of the subdiagram $\gamma_i$. 
Their relation with the numerator of the two-loop diagram $\Gamma$
is specified in \refeq{eq:subdiagnotA}--\refeq{eq:subdiagnotB}.
The required one-loop counterterms and rational terms
associated with the $\gamma_i$ subdiagrams, \ie
$\deltaZ{1}{\gamma_i}{\alpha_i}{q_i}$,
$\deltaZtilde{1}{\gamma_i}{\alpha_i}{\tilde q_i}$, and
$\ratamp{1}{\gamma_i}{\alpha_i}{q_i}$, 
can be found below and in \refapp{sec:UVct}.
Finally, 
the tadpole expansion operators
$\bfSX{i}$ are defined in \refeq{eq:tadexptwoA}--\refeq{eq:tadexptwoC}
(see also \refse{se:tadpoledec}),
and the expansion order $X_i$ is dictated by 
\refeq{eq:Xijdef}--\refeq{eq:Xdef}
and \refeq{eq:powercountingDirectImp}.
After the $\bfSX{i}$ expansions one is left with 
the evaluation of massive tadpole integrals.

In practice, the master formula \refeq{eq:tadexptwoloopL}
and all relevant building blocks have been implemented in 
the {\sc Geficom}~\cite{GEFICOM} framework,
which is based on {\sc Qgraf}~\cite{QGRAF}, {\sc Q2E} and
{\sc Exp}~\cite{Seidensticker:1999bb,Harlander:1997zb} 
for the generation and topology
identification of Feynman diagrams, 
and implements the relevant algebraic manipulations, one-loop insertions
and tadpole decompositions
in {\sc Form}~\cite{Vermaseren:2000nd,Tentyukov:2007mu},
while massive tadpole integrals are computed
with {\sc Matad}~\cite{MATAD}.

We consider the QED Lagrangian
\bea
\mathcal L_{\text{QED}} &=& \bar{\psi}(\ri \gamma^\mu D_\mu - m)\psi 
-\frac{1}{4}F_{\mu\nu}F^{\mu\nu} - \frac{1}{2\,\lambda} (\partial^\mu
A_\mu)^2\,,
\label{eq:lagrangian}
\eea
with $D_\mu = \partial_\mu - i e A_\mu$ and 
a generic gauge parameter $\lambda$.
The corresponding Feynman rules are listed in 
\refapp{sec:UVct} together with the 
known one- and two-loop counterterms in the $\msbar$ scheme.
In the following we present results for the rational terms at one and two
loops in $\dendim=4-2\eps$ dimensions. For convenience we write our results in the
form 
\bea
\label{eq:deltaRdec}
\ratamp{k}{\gamma}{\alpha_1\dots\alpha_N}{} &=& \ri
\lb\frac{\alpha}{4\pi}\rb^k 
S^{k\eps}
\sum_{a} 
\delta \hat\calR^{(a)}_{k,\gamma}\,\,
\calT_{a,\gamma}^{\alpha_1\dots\alpha_N}\,
\,,
\eea
where $k=1,2$ is the loop order, $\alpha=e^2/(4\pi)$, 
$S^\eps$ is the $\msbar$ normalisation factor defined in
\refeq{eq:msbarnorm},
and
$\calT_{a,\gamma}^{\alpha_1\dots\alpha_N}$ are independent tensor structures
carrying the indices $\alpha_1\dots\alpha_N$ of the external lines of the
vertex function at hand.
For convenience the gauge dependence is expressed in terms of
$\gauge=1-\lambda$, \ie the Feynman gauge corresponds to
$\gauge=0$.

The rational terms for the electron two-point function 
have the form 
\begin{align}
&\vcenter{\hbox{\raisebox{0pt}{\includegraphics[width=0.17\textwidth]{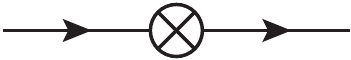}} }}   
&&
\shiftleft
\shiftleft \quad \;\;  = \;
\ri\,
\sum_{k=1}^2 \lb\frac{\alpha}{4\pi}\rb^k 
S^{k\eps}
\bigg[
\delta \hat\calR^{(\srp)}_{k,ee} \,\,
\slashed p_{\alpha\beta} \,
\,+\,
\delta \hat\calR^{(\srm)}_{k,ee}\,\,
m\,\delta_{\alpha\beta}
\bigg]\,,
\label{eq:R2ee}
\end{align}
with two tensor structures, $\calT^{\alpha\beta}_{\srp,ee}=\slashed
p_{\alpha\beta}$,
$\calT^{\alpha\beta}_{\srm,ee}=m\,\delta_{\alpha\beta}$,
and 
the coefficients
\renewcommand{\arraystretch}{1.5}
\begin{align}
\delta \hat \calR_{1,ee}^{(\srp)} &  \;=\;
-1+\frac{2}{3}\gauge\,,             &\quad
\delta \hat \calR_{2,ee}^{(\srp)}   \;=\; &
\left( \frac{19}{18} - \frac{143}{72}\gauge + \frac{11}{30}\gauge^2\right)
{\eps^{-1}}
+ \left(\frac{247}{108} + \frac{293}{864}\gauge + \frac{391}{14400}\gauge^2
\right)\,,
\nonumber\\[2mm] 
\delta \hat \calR_{1,ee}^{(\srm)} & \;=\;   
2-\frac{1}{2}\gauge\,,                  & \quad
\delta \hat \calR_{2,ee}^{(\srm)}   \;=\; &  
\left(-11 + \frac{41}{9}\gauge - \frac{1}{4}\gauge^2 \right) \eps^{-1} 
+ \left( -\frac{5}{6} - \frac{13}{54}\gauge - \frac{7}{288}\gauge^2
\right)\,.
\end{align}
As usual the direction of the momentum $p$ in \refeq{eq:R2ee}
coincides with the fermion flow.

%For the photon two-point function we have
%\begin{align}
%&\vcenter{\hbox{\raisebox{0pt}{\includegraphics[width=0.2\textwidth]{CTphoton}} }} 
%&& \!\!\!\!\!\! = \;
%\ri\,
% \sum_{k=1}^2 \lb \f{\alpha}{4\pi} \rb^k 
%S^{k\eps}
%\bigg[ \delta \hat \calR^{(\srT)}_{k,\gamma \gamma} 
%\lb p^{\mu}p^{\nu} -  p^2 \, g^{\mu \nu} \rb 
%+\lb \delta \hat \calR^{(\srL)}_{k,\gamma \gamma}+
%\delta \tilde{Z}^{(\srL)}_{k,\gamma \gamma}
%\rb \, g^{\mu\nu} 
%\bigg]
%\,, 
%\label{eq:R2aa}
%\end{align}
%with two tensor structures, 
%$\calT^{\mu\nu}_{\srT,\gamma\gamma}\,=\,
%p^{\mu}p^{\nu} -  p^2 \, g^{\mu \nu}$,\,
%$\calT^{\mu\nu}_{\srL,\gamma\gamma}\,=\,
%g^{\mu \nu}$, and
%the coefficients
%
%\begin{align}
%\delta \hat \calR_{1,\gamma\gamma}^{(\srT)}  & \;=\;  
%0\,,           & \quad
%\delta \hat \calR_{2,\gamma\gamma}^{(\srT)}   \;=\; &
% \left(\frac{2}{3}+\frac{4}{9}\gauge \right)\, \eps^{-1} 
%+ \left (- \frac{71}{18} + \frac{59}{108} \gauge
%\right)\,, 
%\nonumber \\[2mm]
%\delta \hat \calR_{1,\gamma\gamma}^{(\srL)} & \;=\;   
%\frac{2}{3} p^2 - 4 m^2\,,              & \quad
%\delta \hat \calR_{2,\gamma\gamma}^{(\srL)}  \;=\;  &  
%  \left( - \frac{11}{12} + \frac{5}{24}\gauge \right)p^2 
%+  \left( 6\, \eps^{-1}+7 \right)m^2\,. 
%\end{align}
%%%%%%%%%%%%%%%%%%%%%%%
For the photon two-point function we have
\begin{align}
&\vcenter{\hbox{\raisebox{0pt}{\includegraphics[width=0.2\textwidth]{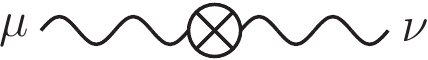}} }} 
&& \!\!\!\!\!\! = \;
\ri\,
 \sum_{k=1}^2 \lb \f{\alpha}{4\pi} \rb^k 
S^{k\eps}
\bigg[ \delta \hat \calR^{(\srp)}_{k,\gamma \gamma} 
\; p^{\mu}p^{\nu} 
+\lb \delta \hat \calR^{(\srG)}_{k,\gamma \gamma} \, p^2 +
\delta \tilde{Z}^{(\srG)}_{k,\gamma \gamma} \, \tilde p^2
\rb \, g^{\mu\nu} 
\bigg]
\,, 
\label{eq:R2aa}
\end{align}
with two tensor structures, 
$\calT^{\mu\nu}_{\srp,\gamma\gamma}\,=\,
p^{\mu}p^{\nu}$,\,
$\calT^{\mu\nu}_{\srG,\gamma\gamma}\,=\,
p^2 g^{\mu \nu}$,
%$\tilde \calT^{\mu\nu}_{\srG,\gamma\gamma}\,=\,
%\tilde p^2 g^{\mu \nu}$, 
and
the coefficients
\begin{align}
\delta \hat \calR_{1,\gamma\gamma}^{(\srp)}  & \;=\;  
0\,,           & \quad
\delta \hat \calR_{2,\gamma\gamma}^{(\srp)}   \;=\; &
 \left(\frac{2}{3}+\frac{4}{9}\gauge \right)\, \eps^{-1} 
+ \left (- \frac{71}{18} + \frac{59}{108} \gauge
\right)\,, 
\nonumber \\[2mm]
\delta \hat \calR_{1,\gamma\gamma}^{(\srG)} & \;=\;   
\frac{2}{3} - 4\,  \f{m^2}{p^2}\,,              & \quad
\delta \hat \calR_{2,\gamma\gamma}^{(\srG)}  \;=\;  &  
\lb -\f{2}{3} -\f{4}{9}\gauge +6\,\f{m^2}{p^2}\rb \eps^{-1}
   + \lb \frac{109}{36} - \frac{73}{216}\gauge +7\,\f{m^2}{p^2}\rb 
\,. 
%\delta \hat \calR_{2,\gamma\gamma}^{(\srG)}  \;=\;  &  
%  \left[ \lb -\f{2}{3} -\f{4}{9}\gauge\rb \eps^{-1}
%   + \lb \frac{109}{36} - \frac{73}{216}\gauge \rb \right] 
%+  \left( 6\, \eps^{-1}+7 \right) \f{m^2}{p^2}\,. 
\end{align}
%%%%%%%%%%%%%%%%%%%%%%%
In addition, due to the presence of a quadratic divergence, the 
usual $\msbar$ counterterm for the 
photon two-point function needs to be supplemented by
\begin{align}
\delta \tilde Z^{(\srG)}_{1,\gamma\gamma}  \;=\;  
\frac{2}{3}
\eps^{-1}\,.
\end{align}
This extra term is relevant only 
when it is inserted in a 
one-loop diagram 
in the context of two-loop 
calculations,
and its two-loop extension 
$\delta \tilde Z^{(\srG)}_{2,\gamma\gamma}$
is required only for calculations beyond two loops.

For the electron-photon vertex
we have
\begin{align}
& \vcenter{\hbox{\raisebox{0pt}{\includegraphics[width=0.17\textwidth]{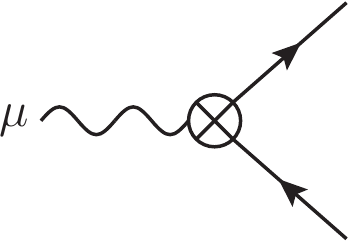}} }} 
&&\shiftleft\shiftleft\shiftleft   = \;
\ri\,e  \gamma^{\mu}\,
\sum_{k=1}^2 \lb \f{\alpha}{4\pi} \rb^k 
S^{k\eps}
\delta \hat \calR^{(\srV)}_{k, ee\gamma}
\,,
\label{eq:R2eea}
\end{align}
with a single tensor structure, $\calT^{\mu}_{\srV,ee\gamma}=e\gamma^\mu$, and
\begin{align}
\delta \hat \calR_{1,ee\gamma}^{(\srV)} &  \;=\;
    -2 + \frac{5}{6}\gauge\,,           & \quad
\delta \hat \calR_{2,ee\gamma}^{(\srV)}   \;=\; &
    \lb \frac{13}{9}- \frac{35}{12}\gauge + \frac{29}{60}\gauge^2 \rb \eps^{-1} + \lb \frac{191}{27}- \frac{17}{24}\gauge + \frac{123}{1600}\gauge^2
\rb\,.
\nonumber\\
\end{align}

Finally, for the quartic photon vertex
we have
\begin{align}
& \vcenter{\hbox{\raisebox{0pt}{\includegraphics[width=0.16\textwidth]{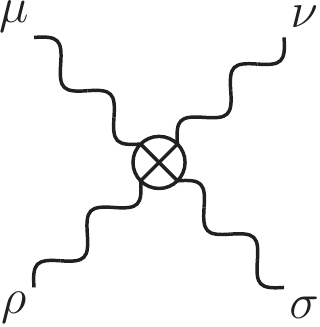} }}} 
&&\shiftleft \;\; = \;
\ri\,
e^2
\, 
\Big(g^{\mu \nu} g^{\rho \sigma} + g^{\mu \rho}g^{\nu \sigma} + g^{\mu \sigma} g^{\nu
\rho}\Big)
\sum_{k=1}^2 \lb \f{\alpha}{4\pi} \rb^k 
S^{k\eps}
\delta \hat
\calR^{(\srS)}_{k,4\gamma} \,,
\label{eq:R24a}
\end{align}
with a single tensor structure, 
$\calT^{\mu\nu\rho\sigma}_{\srS,4\gamma}\,=\,
g^{\mu \nu} g^{\rho \sigma} + g^{\mu \rho}g^{\nu \sigma} + g^{\mu \sigma} g^{\nu
\rho}$, and
\begin{align}
\delta \hat \calR_{1,4\gamma}^{(\srS)} &  \;=\;
    \frac{4}{3}\,,        &\quad
\delta \hat \calR_{2,4\gamma}^{(\srS)}   \;=\; &
    -3 + \frac{1}{2}\gauge\,.
\end{align}

At one loop, the rational counterterms 
$\ratamp{1}{\gamma}{}{}$ are in agreement 
with 
the results obtained in 
\cite{Ossola:2008xq} for $\gauge=0$, 
while their $\gauge$-dependent parts are presented here for the first time.
At two loops, the form of the rational counterterms
$\ratamp{2}{\Gamma}{}{}$ confirms the conclusions of the 
general analysis of 
\refse{se:irredtwoloop}, namely that 
$\ratamp{2}{\Gamma}{}{}$ are polynomials of the external momenta and internal
masses. 
We also note that, due to the presence of $1/\eps^2$ UV poles at two loops,
the $\ratamp{2}{\Gamma}{}{}$ terms contain single $1/\eps$ 
poles. 
Moreover, as expected, the 
$\ratamp{2}{\Gamma}{}{}$ counterterms  are independent of 
the auxiliary tadpole mass $M$.

% and are also free from any 
%logarithms involving the scales of dimensional regularisation and 
%$\msbar$ renormalisation.

\section{Summary and conclusions}

The construction of one-loop scattering amplitudes through efficient
numerical algorithms that handle the numerator of loop integrands in
$\numdim=4$ dimensions turned out to be a very successful strategy for the
automation of NLO calculations.
When the loop numerator is restricted to four dimensions, the
contributions associated with its 
$(\dendim-4)$-dimensional counterpart, referred to as $\ntilde$, 
need to be reconstructed with a different technique.
At one loop, $\ntilde$-contributions can be reconstructed in a very efficient way through the insertion 
of process-independent rational counterterms into tree amplitudes.
In order to open the door to the usage of two-loop numerical algorithms 
in $\numdim=4$ numerator dimensions,
in this paper we have presented a general analysis of 
rational $\ntilde$-contributions at two loops.
Such contributions can arise from the interplay of 
$\ntilde$ with $1/(\dendim-4)$ poles of UV or IR kind, and 
we have focused on poles of UV kind, deferring the 
study of IR poles to future work.

The main result is a formula
that relates generic renormalised two-loop amplitudes with loop numerators
in $\numdim=\dendim$ and $\numdim=4$ dimensions.
Its structure is similar to the well-known \textR-operation for the
subtraction of UV divergences. Renormalised two-loop amplitudes are
expressed as a combination of unrenormalised two-loop amplitudes, one-loop counterterm
insertions into one-loop amplitudes, and two-loop counterterm insertions
into tree amplitudes.
In this formula the well known $\msbar$ counterterms for the subtraction of
UV divergences are accompanied by rational counterterms for the
reconstruction of the related $\ntilde$-contributions.
In addition, the one-loop $\msbar$ counterterms 
for quadratically divergent subdiagrams 
need to be 
supplemented by extra UV counterterms 
proportional to $\tilde
q^2/(\dendim-4)$, where $\tilde q$ 
denotes
the $(\dendim-4)$-dimensional part of the
loop momentum.
The $\ntilde$-contributions associated with one-loop subdivergences are
reconstructed through insertions of the well-known one-loop rational counterterms
into one-loop amplitudes, while the remaining 
$\ntilde$-contributions associated with local two-loop divergences are
reconstructed through the insertion of 
two-loop rational counterterms 
into tree amplitudes.

We have demonstrated that two-loop rational counterterms 
are process-independent polynomials of the external momenta and 
internal masses. They can be extracted from a finite set of 
superficially divergent two-loop diagrams, and 
for their derivation 
we have presented a 
general formula, applicable to any renormalisable theory, 
where the relevant two-loop diagrams are reduced 
to massive tadpole integrals with one auxiliary mass scale, of which the result
is independent.
As a first application we have presented  the full set of two-loop 
rational counterterms for QED in the $R_\xi$-gauge.

\subsection*{Acknowledgements}
We thank J.N.~Lang for numerous discussions. 
H.Z. also thanks A. Primo and T. Peraro for discussions.
This research was supported by the Swiss National Science Foundation (SNSF) 
under contract
BSCGI0-157722, and the work of M.Z. was also supported through the
SNSF Ambizione grant PZ00P2-179877.

\appendix

\section{Feynman rules and UV counterterms in QED}
\label{sec:UVct}
For convenience of the reader  we list 
the Feynman rules for the QED Lagrangian
\refeq{eq:lagrangian} together with the 
full set of $\msbar$ counterterms at one and two loops.
Similarly as in \refse{sec:qedres}, 
the results correspond to 
$\dendim=4-2\eps$ dimensions and are parametrised in terms of
$\eta=1-\lambda$, with $\eta=0$ corresponding to the Feynman gauge.
In analogy with \refeq{eq:deltaRdec}, the UV counterterms are
expressed as 
\bea
\label{eq:uvctstructure}
\deltaZ{k}{\gamma_i}{\alpha_1\dots\alpha_N}{} &=& \ri
\lb\frac{\alpha}{4\pi}\rb^k 
S^{k\eps}
\sum_{a} 
\delta \hat Z^{(a)}_{k,\gamma_i}\,\,
\calT_a^{\alpha_1\dots\alpha_N}\,
\,,
\eea
using the same tensor structures as in \refse{sec:qedres}
and the $\msbar$ normalisation factor~\refeq{eq:msbarnorm}.
The UV counterterms were computed in the same framework as the rational terms and agree with those in the literature, which have been available for a long time
\cite{Jones1974531,PhysRevLett.33.244,Tarasov:1976ef}.

For the electron and photon two-point functions we have
\begin{align}
&\vcenter{\hbox{\raisebox{0pt}{\includegraphics[width=0.17\textwidth]{CTelectron}} }}   
&&
\!\!\!\!\! = \;
\ri\,\Bigg\{
(\slashed p -m)_{\alpha\beta} 
\,+\,
\sum_{k=1}^2 \lb\frac{\alpha}{4\pi}\rb^k 
S^{k\eps}
\bigg[
\delta \hat Z^{(\srp)}_{k,ee} \,\,
\slashed p_{\alpha\beta} \,
\, - \,
\delta \hat Z^{(\srm)}_{k,ee}\,\,
m\,\delta_{\alpha\beta}
\bigg] \Bigg\}\,,
\end{align}
with
\begin{align}
\delta \hat Z_{1,ee}^{(\srp)}  &  \;=\;   
\lb{-1+\gauge}\rb{\eps^{-1}}\,,   &\quad
\delta \hat Z_{2,ee}^{(\srp)}     \;=\; &    
\left(\frac{1}{2}-\gauge+\frac{1}{2}\gauge^2\right){\eps^{-2}}  + \frac{7}{4} {\eps^{-1}}\,,
\nonumber \\[2mm]
\delta \hat Z_{1,ee}^{(\srm)}  &  \;=\;  (-4 + \gauge)\eps^{-1}\,,  & \quad
\delta \hat Z_{2,ee}^{(\srm)}      \;=\;  &     
\left( 6 - 4\gauge + \frac{1}{2}\gauge^2 \right){\eps^{-2}} + \frac{8}{3}{\eps^{-1}}\,, 
\end{align}
and
\begin{align}
&\vcenter{\hbox{\raisebox{0pt}{\includegraphics[width=0.2\textwidth]{CTphoton}} }} 
&& \!\!= \;
-\ri\,
\Bigg\{
%\lb
p^2 \, g^{\mu \nu}
+( \f{1}{\lambda} -1) \, p^{\mu}p^{\nu} 
%x
+ \sum_{k=1}^2 \lb \f{\alpha}{4\pi} \rb^k  
S^{k\eps}
\delta \hat Z^{(\srT)}_{k,\gamma \gamma} \lb p^2 \, g^{\mu \nu}
- p^{\mu}p^{\nu}  \rb \Bigg\}
\,, 
\end{align}
with
\begin{align}
\delta \hat Z_{1,\gamma\gamma}^{(\srT)}  &  \;=\;   
 -\frac{4}{3} \eps^{-1}\,,  & \quad
\delta \hat Z_{2,\gamma\gamma}^{(\srT)}     \;=\; &
 -2\,\eps^{-1}\,.
\end{align}

For the electron--photon vertex 
\begin{align}
& \vcenter{\hbox{\raisebox{0pt}{\includegraphics[width=0.17\textwidth]{CTvertex}} }} 
&&\shiftleft\shiftleft\shiftleft \quad \;\; = \;
\ri\,e  \gamma^{\mu}\, \Bigg\{ 1+
\sum_{k=1}^2 \lb \f{\alpha}{4\pi} \rb^k 
S^{k\eps}
\delta \hat Z^{(\srV)}_{k, ee\gamma}
\Bigg\} \,,
\end{align}
with
\begin{align}
\delta \hat Z_{1,ee\gamma}^{(\srV)}  &  \;=\;   
    (-1+\gauge)\, \eps^{-1}\,,  &\quad
\delta \hat Z_{2,ee\gamma}^{(\srV)}   &  \;=\; 
    \left( \frac{1}{2} - \gauge + \frac{1}{2}\gauge^2 \right) \eps^{-2}  + \frac{7}{4}
\eps^{-1}\,.
\end{align}

In the context of two-loop calculations, when the one-loop counterterms
$\delta Z_{1,\gamma}$ are inserted into one-loop diagrams, the associated tensor
structures and their loop-momentum dependence have to be adapted to the dimensionality of the
loop numerator, \ie using $\delta Z_{1,\gamma}^{\bar \alpha_1\dots \bar
\alpha_N}(\barq_1)$ and $\delta Z_{1,\gamma}^{\alpha_1\dots
\alpha_N}(q_1)$, respectively, in $\numdim=\dendim$ and $\numdim=4$
numerator dimensions.
Moreover, in the master formula \refeq{eq:tadexptwoloopL} the 
four-dimensional $\msbar$ counterterm needs to be supplemented by the additional
$\deltaZtilde{1}{\gamma}{}{\tilde q_1}$
counterterm. The latter is not included in the above formulas since it can
be found in 
\refse{sec:qedres}.

\bibliographystyle{JHEP}
\bibliography{OL_OFRED_literature}

\end{document}

%% file: macrosO2L.tex
\renewcommand{\refeq}[1]{\mbox{\eqref{#1}}}
\newcommand{\refeqs}[2]{\mbox{\eqref{#1}--\eqref{#2}}}
\newcommand{\reffi}[1]{\mbox{Fig.~\ref{#1}}}

\newcommand{\refse}[1]{\mbox{Section~\ref{#1}}}

\newcommand{\refapp}[1]{\mbox{Appendix~\ref{#1}}}

\newcommand{\ie}{i.e.\ }

\newcommand{\f}[2]{\frac{#1}{#2}}
\newcommand{\sss}[1]{\scriptscriptstyle#1}

\newcommand{\nosss}[1]{#1}

\newcommand{\ben}{\begin{enumerate}}
\newcommand{\een}{\end{enumerate}}
\newcommand{\bit}{\begin{itemize}}
\newcommand{\eit}{\end{itemize}}
\newcommand{\bea}{\begin{eqnarray}}
\newcommand{\eea}{\end{eqnarray}}
\newcommand{\be}{\begin{equation}}
\newcommand{\ee}{\end{equation}}
\newcommand{\ba}{\begin{align}}
\newcommand{\ea}{\end{align}}
\newcommand{\beas}{\begin{eqnarray*}}
\newcommand{\eeas}{\end{eqnarray*}}
\newcommand{\bes}{\begin{equation*}}
\newcommand{\ees}{\end{equation*}}
\newcommand{\bas}{\begin{align*}}
\newcommand{\eas}{\end{align*}}

\newcommand{\eps}{{\varepsilon}}

\newcommand{\msbar}{\overline{\mathrm{MS}}}

\newcommand{\lb}{\left(}
\newcommand{\rb}{\right)}

\newcommand{\dendim}{d}
\newcommand{\numdim}{D}
\renewcommand{\dendim}{D}
\renewcommand{\numdim}{D_{\mathrm{n}}}

\newcommand\mseq{\stackrel{\mathclap{\mbox{{\tiny MS}}}}{=}}
\newcommand\msbareq{\stackrel{\mathclap{\overline{\mbox{{\tiny MS}}}}}{=}}

\newcommand{\fmsbar}[1]{f_{\,\overline{\mathrm{{\scriptscriptstyle MS}}}, #1}}
\newcommand{\fms}[1]{f_{\,\mathrm{{\scriptscriptstyle MS}}, #1}}

\newcommand{\denbar}{\bar}

\newcommand{\Dbar}[1]{D_{\nosss{#1}}}

\newcommand{\calA}{\mathcal{A}}
\newcommand{\calC}{\mathcal{C}}
\newcommand{\calD}[1]{\mathcal{D}^{(#1)}(\bar q_{#1})}

\newcommand{\calM}{\mathcal{M}}
\newcommand{\calN}{\mathcal{N}}
\newcommand{\ntilde}{\tilde\calN}

\newcommand{\calR}{\mathcal{R}}
\newcommand{\calT}{\mathcal{T}}

\newcommand{\bfF}{\textbf{F}}
\newcommand{\bfFX}[1]{\bfF^{(#1)}_{X_{#1}}}
\newcommand{\bfK}{\textbf{K}}
\newcommand{\bfKtilde}{\tilde{\textbf{K}}}
\newcommand{\textR}{\textbf{R}}
\newcommand{\bfR}{\textbf{R}}
\newcommand{\bfKsub}{\textbf{K}_{\mathrm{sub}}}
\newcommand{\bfKloc}{\textbf{K}_{\mathrm{loc}}}

\newcommand{\bfKtildesub}{\tilde{\textbf{K}}_{\mathrm{sub}}}
\newcommand{\bfKtildeloc}{\tilde{\textbf{K}}_{\mathrm{loc}}}
\newcommand{\bfS}{\textbf{S}}
\newcommand{\bfSall}{\prod_{i=1}^3\bfS^{(i)}_{X_i}}
\newcommand{\bfSX}[1]{\bfS^{(#1)}_{X_{#1}}}
\newcommand{\barN}{\bar{\mathcal{N}}}

\newcommand{\barA}{\bar{\mathcal{A}}}

\newcommand{\calV}{\mathcal{V}}

\def\prop{\mathrm{prop}}
\def\vert{\mathrm{vert}}
\newcommand{\nprop}[2]{n^{#1,#2}_{\prop}}
\newcommand{\nvert}[2]{n^{#1,#2}_{\vert}}

\newcommand{\singlearg}[1]{
\ifx&#1&%  % #1 is empty
\else
(#1)   % #1 is nonempty
\fi
}

\newcommand{\doublearg}[2]{
\ifx&#2&%  % #2 is empty
(#1)  
\else
(#1,#2)   % #2 is nonempty
\fi
}

\newcommand{\ampindices}[5]{{#1}_{{#2,#3}}^{#4 }\singlearg{#5}}

\newcommand{\amp}[4]{{\ampindices{\calA}{#1}{#2}{#3}{#4}}}
\newcommand{\ampbar}[4]{{\ampindices{\bar\calA}{#1}{#2}{\hspace{0.6pt}#3}{#4}}}
\newcommand{\ratamp}[4]{{\ampindices{\delta \calR}{#1}{#2}{#3}{#4}}}
\newcommand{\deltaZ}[4]{{\ampindices{\delta Z}{#1}{#2}{#3}{#4}}}
\newcommand{\deltaZtilde}[4]{{\ampindices{\delta \tilde Z}{#1}{#2}{#3}{#4}}}
\newcommand{\fullamp}{\bar\calM}

\newcommand{\barq}{\bar q}
\newcommand{\barqidx}[2]{{\bar q}_{#1}^{\hspace{0.9pt}#2}}
\newcommand{\tildeqidx}[2]{{\tilde q}_{#1}^{\hspace{0.6pt}#2}}

\newcommand{\gauge}{\eta}

\def\tad{\mathrm{tad}}

\def\rem{\mathrm{rem}}

\def\baralpham{\bar \alpha}
\def\alpham{\alpha}

\newcommand{\Tr}{\mathrm{Tr}}

\newcommand{\srV}{{\scriptscriptstyle \mathrm V}}
\newcommand{\srT}{{\scriptscriptstyle \mathrm T}}

\newcommand{\srS}{{\scriptscriptstyle \mathrm S}}
\newcommand{\srp}{{\scriptscriptstyle \mathrm P}}
\newcommand{\srG}{{\scriptscriptstyle \mathrm G}}
\newcommand{\srm}{{\scriptscriptstyle \mathrm m}}

\newcommand{\ri}{\mathrm i}
\newcommand{\rd}{\mathrm d}
\newcommand{\ord}{\mathcal O}

\definecolor{bluemar}{rgb}{0,0,.5}
\definecolor{redmar}{rgb}{.8,0,0}
\definecolor{greenmar}{rgb}{0,.5,0}